\documentclass[proof]{pasj01}

\Received{}
\Accepted{}
 
\usepackage{lineno}

  \newcommand{\al}[1]{\begin{eqnarray} #1 \end{eqnarray}}
  \newcommand{\ave}[1]{\left\langle #1 \right\rangle}

\begin{document} 

\title{ 
Generalized local ansatz for scale-dependent primordial non-Gaussianities
and future galaxy surveys }

\author{Daisuke \textsc{Yamauchi}\altaffilmark{1}}
\altaffiltext{1}{Faculty of Engineering, Kanagawa University, Kanagawa, 221-8686, Japan}
\email{yamauchi``at"jindai.jp}

\author{Shuichiro \textsc{Yokoyama},\altaffilmark{2,3}}
\altaffiltext{2}{Kobayashi Maskawa Institute, Nagoya University, Chikusa, Aichi
  464-8602, Japan}
\altaffiltext{3}{Kavli IPMU (WPI), UTIAS, The University of Tokyo, Kashiwa,
  Chiba 277-8583, Japan}
\email{shu``at"kmi.nagoya-u.ac.jp}

\author{Tomo \textsc{Takahashi}\altaffilmark{4}}
\altaffiltext{4}{Department of Physics, Saga University, Saga 840-8502, Japan}
\email{tomot``at"cc.saga-u.ac.jp}


\maketitle

\begin{abstract}
We revisit a possible scale-dependence of local-type primordial non-Gaussianities induced by super-horizon evolution of scalar field perturbations.
We develop the formulation based on $\delta N$ formalism and 
derive the generalized form of the local-type bispectrum and also trispectrum
which allows us to implement the scale-dependence and suitably compare model prediction with observational data.
We propose simple but
phenomenologically meaningful expressions, which encompass the information of a wide range of physically motivated models.
We also formulate large-scale power spectrum and bispectrum of
biased objects in the presence of the scale-dependent primordial non-Gaussianities.
We perform the Fisher analysis for future galaxy surveys and
give the projected 
constraints on the parameters of the generalized local-form of primordial non-Gaussianities.
\end{abstract}


\section{Introduction}

In the standard cosmology, the primordial fluctuations are assumed to be described by a Gaussian distribution, and characterized by an almost scale-invariant power spectrum.
However, 
possible small deviations from a purely Gaussian primordial density fluctuations, called primordial non-Gaussianity, have also been extensively investigated.
Since the primordial non-Gaussianity reflects the fundamental interactions and nonlinear processes involved during and after the inflation,
it can bring insights into the generation mechanism of primordial density fluctuations.
Among several types of primordial non-Gaussianities, the local-type one has been studied widely. 
In the simplest case, the curvature perturbation $\zeta$ can be expanded in terms of the purely Gaussian variable $\zeta_{\rm g}$ as~\citep{Komatsu:2001rj}
\begin{equation}
	\zeta =\zeta_{\rm g}+\frac{3}{5}f_{\rm NL}\zeta_{\rm g}^2+\frac{9}{25}g_{\rm NL}\zeta_{\rm g}^3+\cdots
	\,, \label{eq:simplelocal}
\end{equation}
which leads to the bispectrum and the trispectrum of the following form:
\begin{eqnarray}
	&&B_\zeta (k_1,k_2,k_3)
		=\frac{6}{5}f_{\rm NL}\Bigl[ P_\zeta (k_1)P_\zeta (k_2)+{\rm 2\ perms}\Bigr]
	\,,\\
	&&T_\zeta (k_1k_2,k_3,k_4)
		=\tau_{\rm NL}\Bigl[ P_\zeta (k_1)P_\zeta (k_2)P_\zeta (k_{13})+{\rm 11\ perms}\Bigr]
	\nonumber\\
	&&\quad\quad\quad\quad\quad\quad\quad\quad
			+\frac{54}{25}g_{\rm NL}\Bigl[P_\zeta (k_1)P_\zeta (k_2)P_\zeta (k_3)+{\rm 3\ perms}\Bigr]~.
\end{eqnarray}
Here $f_{\rm NL},~g_{\rm NL},$ and $\tau_{\rm NL}$ are
called nonlinearity parameters for the local-type and usually 
assumed to be constant.
$P_\zeta (k)$ denotes the power spectrum of the curvature perturbations, and $k_{13}=|{\bf k}_1+{\bf k}_3|$.
The generation of the local-type primordial non-Gaussianity through the super-horizon evolution can be evaluated by using the $\delta N$ formalism~\citep{Starobinsky:1986fxa,Salopek:1990jq,Sasaki:1995aw,Sasaki:1998ug,Lyth:2004gb,Lyth:2005fi}.
We consider $\delta N$ with an initial flat hypersurface at $t_{\rm i}$,
on which the initial values of the fields have a certain distribution, and 
follow its evolution
until reaching the final uniform energy density hypersurface at $t_{\rm f}$.
The $\delta N$ formalism gives
the curvature perturbations contributed generally from the multiple scalar fields $\phi^I(t,{\bf x})$ which are 
calculated through the perturbation of the $e$-folding number measured between
the initial and the final surfaces:
\begin{equation}
	\zeta (t_{\rm f},{\bf x})=\delta N(t_{\rm f},\phi^I(t_{\rm i},{\bf x}))
	\,.
\end{equation}
Based on the above formula, expanding in terms of the field perturbations, we can systematically calculate the bispectrum and trispectrum of the primordial curvature perturbations and evaluate the nonlinearity parameters $f_{\rm NL}$, $\tau_{\rm NL}$ and $g_{\rm NL}$.

The nonlinearity parameters have primarily been constrained from the higher-order spectrum in cosmic microwave background (CMB) (see \citep{Akrami:2019izv} for recent constraints from Planck).
However, the current CMB measurements are already reaching 
to the precision
of the cosmic-variance limited one.
A complementary way to access primordial non-Gaussianity is to measure the spatial clustering behavior of halos/galaxies on large scales.
One of the most distinctive effect of the local-type primordial non-Gaussianity is the enhancement of the large-scale clustering of biased objects,
which is due to the non-linear coupling generated by
the primordial non-Gaussianity~\citep{Dalal:2007cu,Desjacques:2008vf}. 
In fact, it has been shown that the power spectrum measurements of galaxy clustering in future surveys such as Square Kilometre Array Observatory (SKAO)~\footnote{
http://www.skatelescope.org}
in radio wavelengths (see e.g., \citep{Bacon:2018dui,Yamauchi:2016ypt}) 
and Euclid~\footnote{
http://www.euclid-ec.org} in optical and infrared bands can constrain the nonlinearity parameters to the level comparable to or tighter 
than CMB measurements (see e.g.~\citep{Ferramacho:2014pua,Yamauchi:2014ioa,dePutter:2014lna,Yamauchi:2015mja}). 
So far, the cosmological analysis of large-scale structure surveys has relied on matter and galaxy power spectrum, however
the extra information can also be extracted with higher-order correlation functions, namely matter and galaxy bispectrum, which is expected 
to have a great potential to drastically improve the constraints on not only the primordial non-Gaussianity~\citep{Jeong:2009vd, Hashimoto:2015tnv,Tellarini:2016sgp, Yamauchi:2016wuc, Shirasaki:2020vkk}
but also modified theories of gravity~\citep{Yamauchi:2017ibz,Yamauchi:2021nxw}.

In this paper, we revisit the extension of the simplest model of the local-type non-Gaussianity
with implementing 
the scale-dependence of the nonlinearity parameters. 
Although in many literatures the local-type nonlinearity parameters are frequently assumed to be scale-independent as in Eq.~(\ref{eq:simplelocal}), 
there are several possible sources that generate a scale-dependent primordial non-Gaussianity.
For instance, in models with multiple scalar fields where a field other than the inflaton can be responsible for the curvature perturbation (see, e.g., \citep{Suyama:2010uj} for various models and their predictions for non-Gaussianities), the non-linear evolution after the horizon exit can generate scale-dependent local-type non-Gaussianities (see e.g., ~\citep{Yokoyama:2007uu,Yokoyama:2007dw,Byrnes:2009pe,Byrnes:2010ft,Byrnes:2012sc} for a general discussion on scale-dependence of non-Gaussianities, see \citep{Byrnes:2010xd,Huang:2010cy,Huang:2010es,Byrnes:2011gh,Huang:2011py,Kobayashi:2012ba,Byrnes:2015asa} for some explicit models).
On the observational side, there have been several works which obtain 
constraints on the scale-dependence of non-Gaussianities from CMB \citep{Becker:2012je,Oppizzi:2017nfy} and large scale structure~\citep{Dai:2019tjh}. Some works investigate expected constraints from future galaxy surveys~\citep{Raccanelli:2014awa,Ballardini:2019wxj} (see also \citep{Sefusatti:2009xu,Giannantonio:2011ya,Becker:2012yr,Biagetti:2013sr} for the pre-Planck analysis). 
Actually,
the galaxy power spectrum analysis alone may not accurately constrain 
the scale-dependence of the primordial non-Gaussianity. 
Hence, we expect that the combination of
the galaxy power and bi-spectra provides us with the unique opportunity to probe the scale-dependence of the primordial non-Gaussianity to some accuracy.  
In the light of this, we revisit the scale-dependence of the nonlinearity parameters which arise in some realistic inflationary models
and explore the impact of the use of both the galaxy power and bi-spectra 
in future large-scale structure surveys.

This paper is organized as follows.
In Sec.~\ref{sec:Generalized local form of primordial non-Gaussianities} we revisit the $\delta N$ formalism to evaluate
the scale-dependence of primordial bi- and tri-spectra.
In particular, we derive the generalized form of the local-type primordial non-Gaussianities at leading order in slow-roll expansion for the scale dependence.
Based on the derived expression, we then calculate the galaxy power- and bi-spectrum, and show the generalized form of 
the scale-dependent biases in Sec.~\ref{sec:Matter clustering with generalized local non-Gaussianities}.
In Sec.~\ref{sec:Results} we quantitatively estimate the impact of the generalized local-type primordial non-Gaussianities and
the use of the galaxy bispectrum to probe primordial non-Gaussianities in SKAO and Euclid as future representative large-scale structure surveys.
Finally, Sec.~\ref{sec:Summary} is devoted to the summary and discussion.

\section{Generalized local form of primordial non-Gaussianities}
\label{sec:Generalized local form of primordial non-Gaussianities}

In this section, we discuss the scale-dependence of the local-type primordial non-Gaussianities generated on super-horizon scales, 
based on the $\delta N$ formalism.
We consider the multiple scalar fields which can contribute 
to a scale-dependence of the nonlinearity parameters. 
As shown below, there are two ways to generate the scale-dependence.
The first one is due to the different scale-dependence of the various Gaussian fields. 
The second one is the evolution of perturbations after horizon exit.
Following Refs.~\citep{Yokoyama:2007uu,Yokoyama:2007dw,Byrnes:2009pe,Byrnes:2010ft}, 
we will show the explicit form of the scale-dependence of 
nonlinearity parameters $f_{\rm NL}, \tau_{\rm NL}$ and $g_{\rm NL}$, in which both of these types are properly incorporated.

\subsection{General argument}

We start with the Lagrangian governing the system, which is given by
\al{
	{\cal L}=-\frac{1}{2}g^{\mu\nu}\delta_{IJ}\partial_\mu\phi^I\partial_\nu\phi^J -V(\phi^K)
	\,.
}
For simplicity, we assume a flat metric in field space.
Based on $\delta N$ formalism~\citep{Starobinsky:1986fxa,Salopek:1990jq,Sasaki:1995aw,Sasaki:1998ug,Lyth:2004gb,Lyth:2005fi}, we expand the Fourier components of the curvature perturbation on super-horizon scales in terms of the fluctuations of the scalar fields as
\al{
	\zeta (t_{\rm f},{\bf k})
		=&&N_I(t_{\rm f},t_{\rm i})\delta\phi^I(t_{\rm i},{\bf k})+\frac{1}{2}N_{IJ}(t_{\rm f},t_{\rm i})\left(\delta\phi^I\star\delta\phi^J\right) (t_{\rm i},{\bf k})
	\nonumber\\
	&&
			+\frac{1}{6}N_{IJK}(t_{\rm f},t_{\rm i})\left(\delta\phi^I\star\delta\phi^J\star\delta\phi^K\right) (t_{\rm i},{\bf k})+\cdots
	\,.\label{eq:deltaN}
}
Here $(\delta\phi\star\delta\phi)$ and $(\delta\phi\star\delta\phi\star\delta\phi)$ denote the convolution defined as
\al{
	\left(\delta\phi^I\star\delta\phi^J\right) (t_{\rm i},{\bf k})
		&& :=\int\frac{\mathrm{d}^3{\bf p}}{(2\pi)^3}
			\delta\phi^I (t_{\rm i},{\bf p})\delta\phi^J(t_{\rm i},{\bf k}-{\bf p})
	\,, \\
\left(\delta\phi^I \star\delta\phi^J \star\delta\phi^K \right)(t_{\rm i},{\bf k}) 
&& := \int\frac{\mathrm{d}^3{\bf p}}{(2\pi)^3} \int\frac{\mathrm{d}^3{\bf q}}{(2\pi)^3} 
\delta\phi^I (t_{\rm i},{\bf p}) \delta\phi^J(t_{\rm i},{\bf q}) \delta\phi^K(t_{\rm i},{\bf k}-{\bf p}-{\bf q}) \,.
}
Here we take a uniform energy density (or comoving) slice at $t_{\rm f}$ and a spatially flat one at $t_{\rm i}$.
The quantities $N(t_{\rm f},t_{\rm i})$, $N_I(t_{\rm f},t_{\rm i}):= \partial N(t_{\rm f},\phi^K (t_{\rm i}))/\partial\phi^I(t_{\rm i})$, 
$N_{IJ}(t_{\rm f},t_{\rm i}):=\partial^2 N(t_{\rm f},\phi^K (t_{\rm i}))/\partial\phi^I(t_{\rm i})\partial\phi^J(t_{\rm i})$ and so on respectively denote 
the number of $e$-folds measured from $t_{\rm i}$ to $t_{\rm f}$, and their derivatives
with respect to scalar fields at $t=t_{\rm i}$.
In the framework of $\delta N$ formalism, the curvature perturbation depends on $t_{\rm f}$, but it should 
be independent on the choice of $t_{\rm i}$.
However, the statistical property of the scalar field perturbations on the flat hypersurface at $t_{\rm i}$ would depend on the choice of $t_i$ due to their super-horizon non-linear growth.
Thus, in usual, $t_{\rm i}$ is taken to be a certain time soon after the relevant length scale crossed the horizon scale, at when the scalar field perturbations on the flat hypersurface are supposed to be almost Gaussian in the slow-roll inflationary phase.
Then the non-Gaussianities of the primordial curvature perturbation can be completely captured in the non-linear evolution of the number of $e$-folds measured from $t_{\rm i}$ to $t_{\rm f}$.
In order to investigate the scale-dependence of 
the bi- and tri-spectra accurately, 
it is quite important to take the fact that the different momenta exit the horizon at different times, very carefully. 

Now let us introduce $t_{\rm c}$ as the time at which all the scales of interest of 
the scalar field fluctuations have already exited the horizon
though the slow-roll conditions for all the relevant scalar fields are still satisfied.
As we have mentioned, the expression (\ref{eq:deltaN}) is independent on the choice of $t_{\rm i}$, and hence $t_{\rm i}$ can be replaced with $t_{\rm c}$. 
In the replaced expression,
$\delta\phi^I(t_{\rm c},{\bf k})$
is no longer the Gaussian due to the super-horizon
non-linear growth.
In order to evaluate the statistical property of $\delta \phi^I(t_{\rm c},{\bf k})$, we need to solve the evolution equations for scalar field perturbations
and obtain the solution for $\delta \phi^I(t_{\rm c},{\bf k})$ order-by-order 
in terms of $\delta\phi^I(t_{\rm i},{\bf k})$ 
(see \citep{Yokoyama:2007uu,Yokoyama:2007dw}).
As far as we consider a mode which exits the horizon before $t=t_{\rm c}$, that is, $k<a(t_{\rm c})H(t_{\rm c})$, 
we can use the slow-roll approximation for solving the super-horizon evolution
and in that approximation the slow-roll parameters are almost constant in time for $t_{\rm i}<t<t_{\rm c}$. Then, we can obtain the formal expressions
as~\citep{Yokoyama:2007uu,Yokoyama:2007dw,Byrnes:2009pe,Byrnes:2010ft}
\al{
	\delta\phi^I (t_{\rm c},{\bf k})
		=&\delta\phi_{(1)}^I(t_{\rm c},{\bf k})
			+\frac{1}{2}\delta\phi_{(2)}^I(t_{\rm c},{\bf k})
			+\frac{1}{6}\delta\phi_{(3)}^I(t_{\rm c},{\bf k})
			+\cdots
	\,,\label{eq:phi_expansion}
}
with
\al{
	&&\delta\phi_{(1)}^I(t_{\rm c},{\bf k})
		=\biggl[\delta^I{}_J+\epsilon^I{}_J\ln\left(\frac{a(t_{\rm c})H(t_{\rm c})}{a(t_{\rm i})H(t_{\rm i})}\right)\biggr]
		\delta\phi^J(t_{\rm i},{\bf k})
		\equiv\Lambda^I{}_J(t_{\rm c},t_{\rm i})\delta\phi^J(t_{\rm i},{\bf k})
	\,,\label{eq:delta phi 1}\\
	&&\delta\phi_{(2)}^I(t_{\rm c},{\bf k})
		=F^I_{(2)JK}\ln\left(\frac{a(t_{\rm c})H(t_{\rm c})}{a(t_{\rm i})H(t_{\rm i})}\right)
			\Bigl(\delta\phi_{(1)}^J\star\delta\phi_{(1)}^K\Bigr) (t_{\rm i},{\bf k})
	\,, \label{eq:delta phi 2} \\
	&&\delta\phi_{(3)}^I(t_{\rm c},{\bf k})
		=F^I_{(3)JKL}\ln\left(\frac{a(t_{\rm c})H(t_{\rm c})}{a(t_{\rm i})H(t_{\rm i})}\right)
			\Bigl(\delta\phi_{(1)}^J\star\delta\phi_{(1)}^K\star\delta\phi_{(1)}^L\Bigr) (t_{\rm i},{\bf k})
	\,.\label{eq:delta phi 3}
}
Here, we have introduced the slow-roll suppressed coefficients $\epsilon^I{}_J$, $F^I_{(2)JK}$, $F^I_{(3)JKL}$,
whose explicit expressions are presented in Appendix \ref{sec:Explicit form of coefficients}.
In order for Eqs.~(\ref{eq:delta phi 1})--(\ref{eq:delta phi 3}) to be valid, 
these parameters are constant in time from $t_{\rm i}$ to $t_{\rm c}$,
and the logarithmic term should not grow very large.
Indeed, when we consider observational scales where the scale-dependence of non-Gaussianities is probed, $\ln(a(t_{\rm c})H(t_{\rm c})/a(t_{\rm i})H(t_{\rm i})) = {\cal O}(1)-{\cal O}(10)$, in which we can safely assume that slow-roll parameters are constant in time. Note that as shown in the above expression for each order, the non-linear relation between $\delta \phi^I(t_{\rm c})$ and $\delta \phi^I (t_{\rm i})$ are expressed at the leading order of $ C_{\rm SR}  \ln(a(t_{\rm c})H(t_{\rm c})/a(t_{\rm i})H(t_{\rm i}))$ with $C_{\rm SR}$ being the coefficient given by the combinations of slow-roll parameters such as $\epsilon^{I}_J$, $F^I_{(2)JK}$, and $F^I_{(3)JKL}$.  As we will see later, this term gives the scale-dependence of the non-linearity parameters.

To evaluate the curvature perturbation, we choose the initial time $t_{\rm i}$ as that at the horizon crossing of a mode ${\bf k}$. Later, we will evaluate the bi- and tri-spectra
of the curvature perturbations which are functions of the multiple wave numbers $\{ {\bf k}_i \}$.
For the different wave numbers, we need to consider the different horizon crossing times. 
Thus, hereinafter we denote the horizon crossing time of the mode $\bf k$ as $t_\ast(k)$ at which $a (t_\ast (k) H (t_\ast(k)) = k$ holds and, in the following, we take $t_{\rm i}$ to be $t_\ast(k)$. Actually the convolution in Eq.~(\ref{eq:delta phi 2}) can be written as
\al{
    \Bigl(\delta\phi_{(1)}^I\star\delta\phi_{(1)}^J\Bigr) (t_{\rm i},{\bf k})
		=\int\frac{\mathrm{d}^3{\bf p}}{(2\pi)^3}
			\delta\phi_{(1)}^I (t_{\rm i},{\bf p})\delta\phi_{(1)}^J(t_{\rm i},{\bf k}-{\bf p})
	\,.
}
(The similar expression holds for the convolution in Eq.~(\ref{eq:delta phi 3}).) Notice that $t_{\rm i}$ is not the horizon crossing time of the modes ${\bf p}$ nor ${\bf k} - {\bf p}$, and hence we need to specify the horizon crossing time for each mode separately.

Based on this formulation, $\delta\phi_{(1)}^I(t,{\bf k})$ can be treated as a Gaussian variable.
Hence, by substituting Eqs.~(\ref{eq:phi_expansion})--(\ref{eq:delta phi 3}) into Eq.~(\ref{eq:deltaN}),
we obtain the curvature perturbation evaluated at $t=t_{\rm c}$ rewritten in terms of
the Gaussian variable $\delta\phi_{(1)}^I(t,{\bf k})$  as
\al{
	\zeta (t_{\rm f},{\bf k})
		=&&N_I(t_{\rm f},t_{\rm c})\delta\phi^I(t_{\rm c},{\bf k})+\frac{1}{2}N_{IJ}(t_{\rm f},t_{\rm c})\left(\delta\phi^I\star\delta\phi^J\right) (t_{\rm c},{\bf k})
	\nonumber\\
	&&\quad\quad\quad\quad
			+\frac{1}{6}N_{IJK}(t_{\rm f},t_{\rm c})\left(\delta\phi^I\star\delta\phi^J\star\delta\phi^K\right) (t_{\rm c},{\bf k})+\cdots
	\nonumber\\
		=&&N_I^\ast (k)\delta\phi_\ast^I({\bf k})
			+\frac{1}{2}N_{IJ}^\ast (k)\Bigl(\delta\phi_{(1)}^I\star\delta\phi_{(1)}^J\Bigr) (t_\ast (k),{\bf k})
	\nonumber\\
	&&\quad\quad\quad\quad
			+\frac{1}{6}N_{IJK}^\ast (k)\Bigl(\delta\phi_{(1)}^I\star\delta\phi_{(1)}^J\star\delta\phi_{(1)}^K\Bigr) (t_\ast (k),{\bf k})
			+\cdots
	\,,\label{eq:zeta expansion}
}
with~\citep{Byrnes:2010ft}
\al{
	&&N_I^\ast (k)=N_J(t_{\rm f},t_{\rm c})\Lambda^J{}_I(t_{\rm c},t_\ast (k))
	\,,\\
	&&N_{IJ}^\ast (k)
		=N_{I^\prime J^\prime}(t_{\rm f},t_{\rm c})\Lambda^{I^\prime}{}_I(t_{\rm c},t_\ast (k))\Lambda^{J^\prime}{}_J(t_{\rm c},t_\ast (k))
	\nonumber\\
	&&\quad\quad\quad\quad\quad
				+N_K(t_{\rm f},t_{\rm c})F^K_{(2)IJ}\ln\left(\frac{a(t_{\rm c})H(t_{\rm c})}{k}\right)
	\,,\\
	&&N_{IJK}^\ast (k)
		=N_{I^\prime J^\prime K^\prime}(t_{\rm f},t_{\rm c})
			\Lambda^{I^\prime}{}_I(t_{\rm c},t_\ast (k))\Lambda^{J^\prime}{}_J(t_{\rm c},t_\ast (k))\Lambda^{K^\prime}{}_K(t_{\rm c},t_\ast (k))
	\nonumber\\
	&&\quad\quad\quad
				+\biggl[
					3N_{I^\prime L}(t_{\rm f},t_{\rm c})\Lambda^{I^\prime}{}_I(t_{\rm c},t_\ast(k) )F^L_{(2)JK}
					+N_{L}(t_{\rm f},t_{\rm c})F^{L}_{(3)IJK}
				\biggr]\ln\left(\frac{a(t_{\rm c})H(t_{\rm c})}{k}\right)
	\,,
}
where for notational simplicity, we denote $\delta\phi_\ast^I ({\bf k}):=\delta\phi^I (t_\ast (k),{\bf k})$.

Based on the derived expressions of the super-horizon evolution of the scalar-field perturbations, 
let us calculate the correlation functions of the curvature perturbation in Fourier space.
The power spectrum of the scalar field perturbations during the slow-roll phase at the horizon crossing time is given by (see Ref.~\citep{Byrnes:2010ft})
\al{ \label{eq:phiPower}
	\ave{\delta\phi_*^I ({\bf k})\delta\phi_*^J({\bf k}^\prime)}
		=&(2\pi)^3\delta^3_{\rm D}({\bf k}+{\bf k})
			\frac{H^2(t_*(k))}{2k^3}\gamma^{IJ}
	\,,
}
where we have introduced $\gamma^{IJ}:=\delta^{IJ}-2\epsilon\delta^I{}_J+2(2-\ln 2-\gamma_{\rm E})\epsilon^I{}_J$
with $\epsilon = - \dot{H}/H^2$ and $\gamma_{\rm E}$ being the Euler-Mascheroni constant.
By using this expression and Eq.~(\ref{eq:zeta expansion}),
we obtain the power spectrum for the curvature perturbation defined by $\ave{\zeta ({\bf k})\zeta({\bf k}^\prime )} =(2\pi )^3\delta^3_{\rm D}({\bf k}+{\bf k}^\prime)P_\zeta (k)$
in the form
\al{ \label{eq:curvPower}
	P_\zeta (k)
		=&\frac{H^2(t_*(k))}{2k^3}N^\ast_I(k)N_\ast^I(k)
	\,,
}
with $N_\ast^I(k):=N^\ast_J(k)\gamma^{IJ}$.
Hereinafter, $t_{\rm f}$ in $\zeta$ is omitted.
This immediately implies that the standard expression for the spectral index
at the leading order as
\al{
	n_{\rm s}-1
		:=\frac{\mathrm{d}\ln (k^3P_\zeta (k))}{\mathrm{d}\ln k}
		=-2\epsilon -2\frac{\epsilon^{IJ}N_I(t_{\rm f},t_{\rm c})N_J(t_{\rm f},t_{\rm c})}{N_K(t_{\rm f},t_{\rm c})N^K(t_{\rm f},t_{\rm c})}
	\,.
}
Actually, in this expression $\epsilon$ term comes from the derivative of $H$ and
$\epsilon^{IJ}$ term can be obtained by the derivative of the linear propagator $\Lambda$
given in Eq.~(\ref{eq:delta phi 1}).

We now proceed to apply this approach to calculate the scale-dependence of the nonlinearity parameters.
By making use of Eq.~(\ref{eq:zeta expansion}) and the fact that
$\delta\phi_{(1)}^I(t,{\bf k})$ obeys the Gaussian statistics,
the three-point function of the curvature perturbation in Fourier space is given by
\al{
	&&\ave{\zeta ({\bf k}_1)\zeta ({\bf k}_2)\zeta ({\bf k}_3)}
		=N_I^\ast (k_1)N_J^\ast (k_2)N^\ast_{KL}(k_3)
	\nonumber\\
	&&\quad\quad\times
			\int\frac{\mathrm{d}^3{\bf p}}{(2\pi)^3}
			\Big\langle\delta\phi_\ast^I({\bf k}_1)\delta\phi_{(1)}^K(t_\ast (k_3),{\bf k}_3-{\bf p})\Big\rangle
			\Big\langle\delta\phi_*^J({\bf k}_2)\delta\phi_{(1)}^L(t_\ast (k_3),{\bf p})\Big\rangle
			+(2~{\rm perms.}) 
	\nonumber\\
	&&\quad
	    =(2\pi)^3\delta_{\rm D}^3({\bf k}_1+{\bf k}_2+{\bf k}_3)
	        N^I_\ast (k_1)N^J_\ast (k_2)N_{KL}^\ast (k_3)
	\nonumber\\
	&&\quad\quad\times
		\Lambda^K{}_I(t_\ast (k_3),t_\ast (k_1))\Lambda^L{}_J(t_\ast (k_3),t_\ast (k_2))
	        \frac{H^2(t_*(k_1))}{2k_1^3}\frac{H^2(t_*(k_2))}{2k_2^3}
	        +(2~{\rm perms.}) 
	\,.
}
Defining the bispectrum as
$\ave{\zeta ({\bf k}_1)\zeta ({\bf k}_2)\zeta ({\bf k}_3)}=(2\pi )^3\delta_{\rm D}^3({\bf k}_1+{\bf k}_2+{\bf k}_3)B_\zeta (k_1,k_2,k_3)$, and using (\ref{eq:curvPower}), we obtain
the expression for the {\it generalized scale-dependent local form} of the bispectrum for the curvature perturbation as~\citep{Byrnes:2010ft}
\al{
	B_\zeta (k_1,k_2,k_3)
		= \frac{6}{5}\widehat f_{\rm NL}(k_1,k_2,k_3)
				P_\zeta (k_1)P_\zeta (k_2)+({\rm 2\ perms.})
	\,,
}
with
\al{
	\frac{6}{5}\widehat f_{\rm NL}(k_1,k_2,k_3)
		=&\frac{N^I_\ast (k_1)N^J_\ast (k_2)N_{KL}^\ast (k_3)\Lambda^K{}_I(t_\ast (k_3),t_\ast (k_1))\Lambda^L{}_J(t_\ast (k_3),t_\ast (k_2))}{N_{I'}^\ast (k_1)N^{I'}_\ast (k_1)N_{J'}^\ast (k_2)N^{J'}_\ast (k_2)}
	\,.\label{eq:fNL def}
}
We have derived this formula by carefully taking account 
of the fact that different modes exit the horizon at different times. 
This expression is generic but a bit complicated. We find that it can be further simplified by expanding it to the first-order
of the expansion.
Keeping up to the first order in $ C_{\rm SR} \ln (a(t_{\rm c})H(t_{\rm c})/k)$, in which  the relation $\Lambda^I{}_J(t_1, t_2) \Lambda^J{}_K(t_2, t_3) \approx  \Lambda^I{}_K(t_1, t_3) $ holds,
we can decompose
 the scale-dependence of $\widehat f_{\rm NL}$ as
\al{
    \frac{6}{5}\widehat f_{\rm NL}(k_1,k_2,k_3)
        \approx\Pi^I(k_1)\Pi^J(k_2)\Theta_{IJ}(k_3)
    \,,\label{eq:reduced fNL formula}
}
where
\al{
   & &\Theta_{IJ}(k):=N_{IJ}(t_{\rm f},t_{\rm c})
				+N_K(t_{\rm f},t_{\rm c})F^K_{(2)IJ}
				\ln\left(\frac{a(t_{\rm c})H(t_{\rm c})}{k}\right)
    \,,\\
    &&\Pi^I(k)
        :=\frac{N^J_\ast (k)\Lambda^I{}_J(t_{\rm c},t_*(k))}{N_K^\ast (k)N^K_\ast (k)}
    \,.\label{eq:Pi def}
}
This simplified expression is one of the main results in this paper.
We found that the resultant formula of $\widehat f_{\rm NL}$ can be described by 
a product of the functions depending on a single $k$-mode. Actually the form of Eq.~(\ref{eq:reduced fNL formula}) has been phenomenologically introduced to 
study its effects on the halo bias in \cite{Shandera:2010ei}, 
however, we have derived this form from the $\delta N$ formalism 
in a general setup and given the explicit formulas for $\Theta_{IJ}$ 
and $\Pi^I$ as a function of $k$.

Following the same step as the three-point function, 
the four-point function of the curvature perturbation in Fourier space is given as
\al{
	&&\ave{\zeta ({\bf k}_1)\zeta ({\bf k}_2)\zeta ({\bf k}_3)\zeta ({\bf k}_4)}
		=N_I^\ast (k_1)N^\ast_J(k_2)N^\ast_{KL}(k_3)N^\ast_{MN}(k_4)
	\nonumber\\
	&&\quad\quad\times
			\int\frac{\mathrm{d}^3{\bf p}_1\mathrm{d}^3{\bf p}_2}{(2\pi)^6}
			\Big\langle\delta\phi_\ast^I({\bf k}_1)\delta\phi_{(1)}^K(t_\ast (k_3),{\bf k}_3-{\bf p}_1)\Big\rangle
    \nonumber\\
    &&\quad\quad\times	
				\Big\langle\delta\phi_\ast^J({\bf k}_2)\delta\phi_{(1)}^M(t_\ast (k_4),{\bf k}_4-{\bf p}_2)\Big\rangle
				\Big\langle\delta\phi_{(1)}^L(t_\ast (k_3),{\bf p}_1)\delta\phi_{(1)}^N(t_\ast (k_4),{\bf p}_2)\Big\rangle
	\nonumber\\
	&&\quad
			+(11~{\rm perms.})
	\nonumber\\
	&&\quad
		+N^\ast_I(k_1)N^\ast_J(k_2)N^\ast_K(k_3)N^\ast_{LMN}(k_4)
	\nonumber\\
	&&\quad\quad\times
		\int\frac{\mathrm{d}^3{\bf p}_1\mathrm{d}^3{\bf p}_2}{(2\pi)^6}
			\Big\langle\delta\phi_*^I({\bf k}_1)\delta\phi_{(1)}^L(t_\ast (k_4),{\bf k}_4-{\bf p}_1-{\bf p}_2)\Big\rangle
	\nonumber\\
	&&\quad\quad\times	
				\Big\langle\delta\phi_*^J({\bf k}_2)\delta\phi_{(1)}^M(t_\ast (k_4),{\bf p}_1)\Big\rangle
				\Big\langle\delta\phi_*^K({\bf k}_3)\delta\phi_{(1)}^N(t_\ast (k_4),{\bf p}_2)\Big\rangle
			+(3~{\rm perms.})
	\nonumber\\
	&&\quad
	    =(2\pi)^3\delta_{\rm D}^3({\bf k}_1+{\bf k}_2+{\bf k}_3+{\bf k}_4)
	\nonumber\\
	&&\quad\quad\times
	        \Biggl\{
	            \biggl[
	                N^I_\ast (k_1)N_\ast^J(k_2)
			        N^\ast_{KL}(k_3)
			        N^\ast_{MN}(k_4)
				\Lambda^K{}_I(t_\ast (k_3),t_\ast (k_1))\Lambda^M{}_J(t_\ast (k_4),t_\ast (k_2))
	\nonumber\\
	&&\quad\quad\quad\quad\quad\times
				\Lambda^L{}_{L^\prime}(t_\ast (k_3),t_\ast (k_{13}))\Lambda^N{}_{N^\prime}(t_\ast (k_4),t_\ast (k_{13}))
			        \gamma^{L^\prime N^\prime}
	\nonumber\\
	&&\quad\quad\quad\quad\quad\times
			         \frac{H^2(t_*(k_1))}{2k_1^3}\frac{H^2(t_*(k_2))}{2k_2^3}
			          \frac{H^2(t_*(k_{13}))}{2k_{13}^3}
			          +(11~{\rm perms.})
			    \biggr]
    \nonumber\\
    &&\quad\quad\quad
                +\biggl[
                    N_\ast^I(k_1)N_\ast^J(k_2)N_\ast^K(k_3)N^{\ast}_{LMN}(k_4)
	\nonumber\\
	&&\quad\quad\quad\quad\quad\times
			\Lambda^{L}{}_I(t_\ast (k_4),t_\ast (k_1))\Lambda^M{}_J(t_\ast (k_4),t_\ast (k_2))\Lambda^N{}_K(t_\ast (k_4),t_\ast (k_3))
  	\nonumber\\
	&&\quad\quad\quad\quad\quad\times
                    \frac{H^2(t_*(k_1))}{2k_1^3}\frac{H^2(t_*(k_2))}{2k_2^3}
			        \frac{H^2(t_*(k_3))}{2k_3^3}
			          +(3~{\rm perms.})
                \biggr]
			\Biggr\} \,.
}
Defining the trispectrum for the curvature perturbation as 
$\ave{\zeta ({\bf k}_1)\zeta ({\bf k}_2)\zeta ({\bf k}_3)\zeta ({\bf k}_4)}=(2\pi )^3\delta_{\rm D}^3({\bf k}_1+{\bf k}_2+{\bf k}_3+{\bf k}_4)~ T_\zeta (k_1,k_2,k_3,k_4)$, 
we write down the {\it generalized scale-dependent local form} of 
the trispectrum for the curvature perturbation as~\citep{Byrnes:2010ft}
\al{
    T_\zeta (k_1k_2,k_3,k_4)
		=&&\widehat\tau_{\rm NL}(k_1,k_2,k_3,k_4)P_\zeta (k_1)P_\zeta (k_2)P_\zeta (k_{13})+({\rm11\ perms.})
		\nonumber\\
		&& +\frac{54}{25}\widehat g_{\rm NL}(k_1,k_2,k_3,k_4)P_\zeta (k_1)P_\zeta (k_2)P_\zeta (k_3)+({\rm 3\ perms.})
    \,,
}
with
\al{
	&&\widehat\tau_{\rm NL}(k_1,k_2,k_3,k_4)
		= \frac{N^I_\ast (k_1)N_\ast^J(t_\ast (k_2))N^\ast_{KL}(k_3)N^\ast_{MN}(k_4)}{N_{I^\prime}(k_1)N^{I^\prime}(k_1)N_{J^\prime}(k_2)N^{J^\prime}(k_2)N_{K^\prime}(k_{13})N^{K^\prime}(k_{13})}
	\nonumber\\
	&&\quad\quad\quad\quad\quad\quad\quad\times
				\Lambda^K{}_I(t_\ast (k_3),t_\ast (k_1))\Lambda^M{}_J(t_\ast (k_4),t_\ast (k_2))
	\nonumber\\
	&&\quad\quad\quad\quad\quad\quad\quad\times
				\Lambda^L{}_{L^\prime}(t_\ast (k_3),t_\ast (k_{13}))\Lambda^N{}_{N^\prime}(t_\ast (k_4),t_\ast (k_{13}))
			        \gamma^{L^\prime N^\prime}
	\,,\label{eq:tauNL sol}\\
	&&\frac{54}{25}\widehat g_{\rm NL}(k_1,k_2,k_3,k_4)
		=\frac{N_\ast^I(k_1)N_\ast^J(k_2)N_\ast^K(k_3)N^{\ast}_{LMN}(k_4)}{N_{I^\prime}^{\ast}(k_1)N^{I^\prime}_{\ast}(k_1)N_{J^\prime}^{\ast}(k_2)N^{J^\prime}_{\ast}(k_2)N_{K^\prime}^{\ast}(k_3)N^{K^\prime}_{\ast}(k_3)}
	\nonumber\\
	&&\quad\quad\quad\quad\quad\quad\quad\times
			\Lambda^{L}{}_I(t_\ast (k_4),t_\ast (k_1))\Lambda^M{}_J(t_\ast (k_4),t_\ast (k_2))\Lambda^N{}_K(t_\ast (k_4),t_\ast (k_3))
	\,.
}
We expand these expression with respect to $C_{\rm SR} \ln (a(t_{\rm c})H(t_{\rm c})/k)$ and keep the first order term. 
We then obtain
\al{
    &&\widehat\tau_{\rm NL}(k_1,k_2,k_3,k_4)
        \approx\Pi^I(k_1)\Pi^J(k_2)
            \Theta_{IK}(k_3)\Theta_{JL}(k_4) \Gamma^{KL}(k_{13})
    \,,\label{eq:tauNL def}\\
    &&\frac{54}{25}\widehat g_{\rm NL}(k_1,k_2,k_3,k_4)
        \approx\Pi^I(k_1)\Pi^J(k_2)\Pi^K(k_3)\Theta_{IJK}(k_4)
    \,,\label{eq:gNL def}
}
where
\al{
    &&\Gamma^{IJ}(k)
        :=\frac{\Lambda^I{}_{K}(t_{\rm c},t_\ast (k))
            \Lambda^J{}_{L}(t_{\rm c},t_\ast (k))\gamma^{KL}}{N_M^\ast(k)N_\ast^M(k)}
    \,,\label{eq:Gamma def}\\
	&&\Theta_{IJK}(k)
	:=N_{IJK}(t_{\rm f},t_{\rm c})
		+\Bigl[
			3N_{IL}(t_{\rm f},t_{\rm c})F^L_{(2)JK}
			+N_{L}(t_{\rm f},t_{\rm c})F^{L}_{(3)IJK}
		\Bigr]\ln\left(\frac{a(t_{\rm c})H(t_{\rm c})}{k}\right)
	\,.\label{eq:Theta_IJK def}
}
Similar to $\widehat f_{\rm NL}$, we have shown that $\widehat\tau_{\rm NL}$ and $\widehat g_{\rm NL}$ 
can be also written as a product of the functions
of a single $k$ mode, which is also one of our main results in this paper.
Thus, we have derived the generalized scale-dependent form of the local-type bispectrum and trispectrum based on $\delta N$ formalism.
The resultant formulae show that 
the scale-dependence of
the nonlinearity parameters can be induced from
the evolution of the long modes after horizon crossing, which are characterized
by $\epsilon^{IJ}, F_{(2)JK}^I$, and $F_{(3)JKL}^I$.
Moreover, the nonlinearity parameters would have the scale-dependence due to 
the contributions from multi-scalar fields which have the different scale dependence in
the power spectrum at the horizon crossing time.

\subsection{Parameterizing scale-dependent local form of primordial non-Gaussianities}

Although the formulae derived in the previous subsection are generic,
they seem complicated to estimate the constraints from the observations.
Here, based on the above generic expressions, we propose a simpler but phenomenologically meaningful expression for the generalized scale-dependent local form of the primordial non-Gaussianities.
To do this, we introduce two functions to describe the scale-dependent nonlinearity parameter for primordial bispectrum:
(see also \citep{Shandera:2010ei})
\al{
	\widehat f_{\rm NL}(k_1,k_2,k_3)
		=f_{\rm NL}\xi_{\rm m}(k_1)\xi_{\rm m}(k_2)\xi_{\rm s}(k_3)
	\,.\label{eq:generalized local bis}
}
This form captures a wide range of physically motivated models.
As shown below, two functions $\xi_{\rm s}(k)$ and $\xi_{\rm m}(k)$ are related with 
the non-linear super-horizon evolution of the scalar field perturbations included in $\Theta_{IJ}(k)$
and the multi-field contributions in the curvature perturbation
characterized by $\Pi^I(k)$, respectively.
Based on the analysis presented in the previous subsections, in order to describe the scale-dependence of the primordial trispectrum, 
we need to add 
a function $\xi_{\rm g}(k)$ characterized by $\Theta_{IJK}(k)$. 
We now define the scale-dependent functions as the generalized local form of
the primordial trispectrum:
\al{
	&&\widehat\tau_{\rm NL}(k_1,k_2,k_3,k_4)
		=\tau_{\rm NL}\xi_{\rm m}(k_1)\xi_{\rm m}(k_2)\xi_{\rm m}(k_{13})\xi_{\rm s}(k_3)\xi_{\rm s}(k_4)
	\,,\label{eq:generalized local tauNL}\\
	&&\widehat g_{\rm NL}(k_1,k_2,k_3,k_4)
		=g_{\rm NL}\xi_{\rm m}(k_1)\xi_{\rm m}(k_2)\xi_{\rm m}(k_3)\xi_{\rm g}(k_4)
	\,.\label{eq:generalized local tri}
}
In the subsequent analysis, we will parametrize the scale-dependence as
\al{
	\xi_{\rm A}(k)=\left(\frac{k}{k_{\rm piv}}\right)^{n_{\rm NL}^{\rm A}}
	\,,\label{eq:xi def}
}
where ${\rm A}={\rm s},{\rm m}$, and ${\rm g}$, and $k_{\rm piv}$ is a pivot scale. 
Hereafter, we will show several models in which the scale-dependence of primordial non-Gaussianities
is naturally induced and their correspondence to the above formula.

\subsubsection{Generic single-source case}
\label{sec:Generic single-source case}

Let us consider the simplest example where the primordial curvature perturbation is generated from a single source, say $\phi$. Examples of this case includes self-interacting curvaton and axionic curvaton models where the curvaton alone generates the curvature perturbations \citep{Byrnes:2009pe,Byrnes:2010xd,Huang:2010cy,Byrnes:2011gh,Byrnes:2015asa}.
In this case, the resultant formulae derived in the previous subsection is drastically simplified.
In particular, $\Pi^\phi$ and $\Gamma^{\phi\phi}$ defined in Eqs.~(\ref{eq:Pi def}) and (\ref{eq:Gamma def}) can be reduced to the simple forms:
\al{
    &&\Pi^\phi(k)
        =\frac{\Lambda^\phi{}_\phi (t_{\rm c},t_\ast(k))}{N_\phi^\ast (k)}
        =\frac{1}{N_\phi (t_{\rm f},t_{\rm c})}
    \,,\\
    &&\Gamma^{\phi\phi}(k)
        =\left(\frac{\Lambda^\phi{}_\phi (t_{\rm c},t_\ast(k))}{N_\phi^\ast (k)}\right)^2
        =\frac{1}{N_\phi^2 (t_{\rm f},t_{\rm c})}
    \,,
}
where we have used the relation Eq.~(\ref{eq:delta phi 1}).
An interesting observation is that the scale-dependence of $\Pi^\phi$ and $\Gamma^{\phi\phi}$ completely vanishes in the single-source case.
Hence, substituting these into Eqs.~(\ref{eq:fNL def}), (\ref{eq:tauNL def}), and (\ref{eq:gNL def}),
we show that the leading-order scale-dependence of the nonlinearity parameters
can be written in the following forms:
\al{
    &&\frac{6}{5}\widehat f_{\rm NL}
		=\frac{N_{\phi\phi}(t_{\rm f},t_{\rm c})}{N_\phi^2(t_{\rm f},t_{\rm c})} \left[1 + \frac{N_{\phi}(t_{\rm f},t_{\rm c})F_{(2)\phi\phi}^\phi}{N_{\phi\phi}(t_{\rm f},t_{\rm c})} \ln \left(\frac{a(t_{\rm c})H(t_{\rm c})}{k_3} \right)\right]
	\,,\\
	&&\widehat\tau_{\rm NL}
		=\frac{N_{\phi\phi}^2(t_{\rm f},t_{\rm c})}{N_\phi^4(t_{\rm f},t_{\rm c})}\left[1 + \frac{N_{\phi}(t_{\rm f},t_{\rm c})F_{(2)\phi\phi}^\phi}{N_{\phi\phi}(t_{\rm f},t_{\rm c})} 
		\left( \ln \left(\frac{a(t_{\rm c})H(t_{\rm c})}{k_3} \right) 
		+
		\ln \left(\frac{a(t_{\rm c})H(t_{\rm c})}{k_4} \right) 
		\right)
		\right]
	\,,\\
	&&\frac{54}{25}\widehat g_{\rm NL}
		=\frac{N_{\phi\phi\phi}(t_{\rm f},t_{\rm c})}{N_\phi^3(t_{\rm f},t_{\rm c})} \left[ 1 +\frac{3N_{\phi\phi}(t_{\rm f},t_{\rm c})F_{(2)\phi\phi}^\phi +N_{\phi}(t_{\rm f},t_{\rm c})F_{(3)\phi\phi\phi}^\phi}{N_{\phi\phi\phi}(t_{\rm f},t_{\rm c})}\ln \left(\frac{a(t_{\rm c})H(t_{\rm c})}{k_4} \right)\right]
	\,.
}
Thus, comparing the above result with Eqs.~(\ref{eq:generalized local bis}), (\ref{eq:generalized local tauNL}) and (\ref{eq:generalized local tri}), the nonlinearity parameters in the single-source case do not have the scale-dependence characterized by $\xi_{\rm m}(k)$.
By assuming the coefficient in front of $\ln k$ is small, for $f_{\rm NL}$ we can approximately obtain
\al{
    &&\ln\left(\frac{6}{5}\widehat f_{\rm NL}\right)= \ln \left\{\frac{N_{\phi\phi}(t_{\rm f},t_{\rm c})}{N_\phi^2(t_{\rm f},t_{\rm c})} \left[1 + \frac{N_{\phi}(t_{\rm f},t_{\rm c})F_{(2)\phi\phi}^\phi}{N_{\phi\phi}(t_{\rm f},t_{\rm c})} \ln \left(\frac{a(t_{\rm c})H(t_{\rm c})}{k_3} \right)\right] \right\}
		\nonumber\\
    && \qquad \qquad 
    = \ln \left( \frac{N_{\phi\phi}(t_{\rm f},t_{\rm c})}{N_\phi^2(t_{\rm f},t_{\rm c})}  \right) + \ln \left[1 + \frac{N_{\phi}(t_{\rm f},t_{\rm c})F_{(2)\phi\phi}^\phi}{N_{\phi\phi}(t_{\rm f},t_{\rm c})} \ln \left(\frac{a(t_{\rm c})H(t_{\rm c})}{k_3} \right)\right]
    	\nonumber\\
    && \qquad \qquad
    \simeq  \ln \left( \frac{N_{\phi\phi}(t_{\rm f},t_{\rm c})}{N_\phi^2(t_{\rm f},t_{\rm c})}  \right) + \frac{N_{\phi}(t_{\rm f},t_{\rm c})F_{(2)\phi\phi}^\phi}{N_{\phi\phi}(t_{\rm f},t_{\rm c})} \ln \left(\frac{a(t_{\rm c})H(t_{\rm c})}{k_3} \right)
    	\nonumber\\
    && \qquad \qquad
    =  \ln \left[ \frac{N_{\phi\phi}(t_{\rm f},t_{\rm c})}{N_\phi^2(t_{\rm f},t_{\rm c})} \left(\frac{k_3}{a(t_{\rm c})H(t_{\rm c})} \right)^{n_{\rm NL}^{\rm s}} \right]~,
}
with 
\al{
    &n_{\rm NL}^{\rm s}=-\frac{N_\phi(t_{\rm f},t_{\rm c})}{N_{\phi\phi}(t_{\rm f},t_{\rm c})}F^\phi_{(2)\phi\phi}
    \,.
}
Thus, in the single-source case the scale-dependent
local-type $\widehat f_{\rm NL}$ can be leadingly
parameterized as Eq.~(\ref{eq:generalized local bis}) 
with Eq.~(\ref{eq:xi def}).

Similarly, the scale-dependence of the local-type $\widehat \tau_{\rm NL}$ and $\widehat g_{\rm NL}$
are approximately described as
\al{
&&\widehat \tau_{\rm NL} \simeq \left( \frac{N_{\phi\phi}(t_{\rm f},t_{\rm c})}{N_\phi^2(t_{\rm f},t_{\rm c})} \right)^2 \left(\frac{k_3}{a(t_{\rm c})H(t_{\rm c})} \right)^{n_{\rm NL}^{\rm s}}
\left(\frac{k_4}{a(t_{\rm c})H(t_{\rm c})} \right)^{n_{\rm NL}^{\rm s}}
	\,,\\
&&\frac{54}{25} \widehat g_{\rm NL} \simeq 
\left( \frac{N_{\phi\phi\phi}(t_{\rm f},t_{\rm c})}{N_\phi^3(t_{\rm f},t_{\rm c})} \right)
\left(\frac{k_3}{a(t_{\rm c})H(t_{\rm c})} \right)^{n_{\rm NL}^{\rm g}} \,,
}
with 
\al{
    n_{\rm NL}^{\rm g}=-\frac{1}{N_{\phi\phi\phi}(t_{\rm f},t_{\rm c})}
            \biggl[ 3N_{\phi\phi}(t_{\rm f},t_{\rm c})F^\phi_{(2)\phi\phi}+N_\phi(t_{\rm f},t_{\rm c}) F^\phi_{(3)\phi\phi\phi}\biggr]
    \,.
}

\subsubsection{Mixed inflaton-spectator scenario}
\label{sec:Mixed inflation-curvaton scenario}

Next, as a specific model of multi-field models, we consider the model 
of an inflaton field $\sigma$ and a spectator
field $\phi$.
We now assume that the primordial curvature perturbation generated from
the inflaton fluctuation is very close to Gaussian,
while the primordial curvature perturbation from 
the spectator fluctuation 
may provide significantly large non-Gaussianity. 
Thus, the primordial curvature perturbation can be well approximated as
\al{
	\zeta ({\bf k})=&&N_\sigma (t_{\rm f},t_{\rm i})\delta\sigma (t_{\rm i},{\bf k})+N_\phi (t_{\rm f},t_{\rm i})\delta\phi (t_{\rm i},{\bf k})
		+\frac{1}{2}N_{\phi\phi}(t_{\rm f},t_{\rm i})\left(\delta\phi\star\delta\phi\right) (t_{\rm i},{\bf k})
	\nonumber\\
	&&
		+\frac{1}{6}N_{\phi\phi\phi}(t_{\rm f},t_{\rm i})\left(\delta\phi\star\delta\phi\star\delta\phi\right) (t_{\rm i},{\bf k})
		+\cdots
	\,.
}
Some explicit models of this type have been discussed in \citep{Byrnes:2009pe,Byrnes:2010ft,Kobayashi:2012ba}.
To simplify the analysis, we further assume that the cross term $\epsilon_{\phi\sigma}$ is negligibly smaller than other components $\epsilon_{\phi\phi}$ and $\epsilon_{\sigma\sigma}$,
which immediately leads to $\ave{\delta\phi_\ast ({\bf k}_1)\delta\sigma_\ast ({\bf k}_2)}=0$.
Under these assumptions, one sees that the power spectrum for the primordial curvature perturbation can be decomposed into two parts:
\al{
	P_\zeta (k)
	= \frac{H^2(t_*(k))}{2k^3}\Bigl[         
            (N_\sigma^\ast (k))^2
	        \gamma^{\sigma\sigma}
	        + (N_\phi^\ast (k))^2
	        \gamma^{\phi\phi}
	        \Bigr]
		\equiv P_{\zeta ,\sigma}(k)+P_{\zeta ,\phi}(k)
	\,.
}
Even in this case, since the non-Gaussianity comes from the nonlinear terms such as $N_{\phi \phi}$ and $N_{\phi \phi \phi}$, the nonlinearity parameters are given by
\al{
	&&\frac{6}{5}\widehat f_{\rm NL}(k_1,k_2, k_3)
		=\Pi^\phi(k_1)\Pi^\phi(k_2)\Theta_{\phi\phi}(k_3)
	\,,\label{eq:fNL multi} \\
	&&\widehat\tau_{\rm NL}(k_1,k_2,k_3,k_4)
        =\Pi^\phi(k_1)\Pi^\phi(k_2)
            \Theta_{\phi\phi}(k_3)\Theta_{\phi\phi}(k_4)
            \Gamma^{\phi\phi} (k_{13})
    \,,\label{eq:tauNL multi}\\
    &&\frac{54}{25}\widehat g_{\rm NL}(k_1,k_2,k_3,k_4)
        =\Pi^\phi(k_1)\Pi^\phi(k_2)\Pi^\phi(k_3)\Theta_{\phi\phi\phi}(k_4)
    \,,\label{eq:gNL multi}
}
where $\Pi^\phi$ and $\Gamma^{\phi\phi}$ can be written as
\al{
	&&\Pi^\phi (k)
	    =\frac{R(k)}{1+R(k)}\frac{1}{N_\phi (t_{\rm f},t_{\rm c})}
	\,,\\
	&&\Gamma^{\phi\phi}(k)
	    =\frac{R(k)}{1+R(k)}\frac{1}{N_\phi^2 (t_{\rm f},t_{\rm c})}
	  \,.
}
Here we have used Eq.~(\ref{eq:delta phi 1}) and we have introduced the fractional power spectrum defined as
\al{
	R(k)=\frac{P_{\zeta, \phi}(k)}{P_{\zeta ,\sigma} (k)}
	\,.
}
Comparing with the single-source case discussed previously, in the mixed scenario we have additional scale-dependence characterized by $R(k)$, which is induced from
the different
scale-dependence of $P_{\zeta,\phi}$ and $P_{\zeta,\sigma}$.
In the limit where $R \gg 1$ for all the relevant scales, the above expressions for the nonlinearity parameters actually come to those in the single-source case.
Thus, the additional scale-dependence
in the mixed scenario 
can be captured in the function, $\xi_{\rm m}(k) \propto k^{n_{\rm NL}^{\rm m}}$
with the power-law index:
\al{
n_{\rm NL}^{\rm m} = \frac{2}{1+R} \left( \epsilon^{\sigma}{}_{\sigma} - \epsilon^\phi{}_\phi \right)~.
}

\section{Matter clustering with generalized local non-Gaussianities}
\label{sec:Matter clustering with generalized local non-Gaussianities}

In this section, we calculate the power spectrum and bispectrum of biased objects, based on
the integrated perturbation theory (iPT)~\citep{Matsubara:2011ck}. 
Provided the statistical nature of primordial curvature perturbation, the linear density field
$\delta_{\rm L}$ is determined through
\al{
	\delta_{\rm L}({\bf k},z)={\cal M}(k;z)\Phi ({\bf k})
	\,,\label{eq:linear density field}
}
where the function ${\cal M}(k,z)$ are defined as
\al{
	{\cal M}(k;z)=\frac{2}{3}\frac{D_+(z)}{D_+(z_*)(1+z_*)}\frac{k^2T(k)}{H_0^2\Omega_{{\rm m},0}}
	\,,
}
with
$D_+(z)$ and $T(k)$ 
being
the linear growth rate and matter transfer function, respectively.
$z_*$ represents an arbitrary redshift at the matter dominated era.
With these, we also introduce the power and higher-order spectra of density field, which are directly related to those of
the primordial spectra as 
$P_{\rm L}(k)={\cal M}^2(k)P_\Phi (k)$, $B_{\rm L}({\bf k}_1,{\bf k}_2,{\bf k}_3)={\cal M}(k_1){\cal M}(k_2){\cal M}(k_3)B_\Phi ({\bf k}_1,{\bf k}_2,{\bf k}_3)$
and $T_{\rm L}({\bf k}_1,{\bf k}_2,{\bf k}_3,{\bf k}_4)={\cal M}(k_1){\cal M}(k_2){\cal M}(k_3){\cal M}(k_4)T_\Phi ({\bf k}_1,{\bf k}_2,{\bf k}_3,{\bf k}_4)$
for the linear density field.
The variation of density fluctuations smoothed on a scale $R$ associated to mass $M$ is defined as
\al{
	\sigma^2(M)=\int\frac{\mathrm{d}^3{\bf k}}{(2\pi)^3}W^2(kR)P_{\rm L}(k)
	\,,
}
where $W(kR)$ is the Fourier transform of the top-hat window function. The spatial smoothing scale $R$ is the related to
the smoothing mass scale $M$ via $M=(4\pi /3)R^3\rho_{{\rm m},0}$, 
with $\rho_{{\rm m},0}$ being the matter energy density at the present time.

In this paper, we focus on the clustering features of galaxy number density field obtained from future galaxy surveys.
Although the linear density field is indirectly related to observables of large-scale structure,
the relation between them is rather complicated. 
To connect them, we will adopt the iPT~\citep{Matsubara:2011ck,Matsubara:2012nc} to include the late-time gravitational evolution and 
the effect of halo/galaxy bias without assuming the peak-background split or the high-peak limit.
The explicit relation between the linear density field and
the galaxy power spectrum and bispectrum in the iPT framework
is presented in Appendix \ref{sec:Integrated perturbation theory}.

\subsection{Power spectrum of biased object in iPT}
\label{sec:Galaxy power spectrum}

The power spectrum of the biased objects ${\rm X}$ and ${\rm Y}$ up to the leading order of
the nonlinearity parameters can be represented as~\citep{Yokoyama:2012az}
\al{
	P_{\rm XY}(k)
		=P_{\rm XY}^{\rm tree}(k) +P_{\rm XY}^{\rm bis}(k)+P_{\rm XY}^{\rm tri}(k)+\cdots
	\,,
}
where
\al{
	&&P_{\rm XY}^{\rm tree}(k)
		=\Gamma_{\rm X}^{(1)}({\bf k})\Gamma_{\rm Y}^{(1)}({\bf k})P_{\rm L}(k)
	\,,\\
	&&P_{\rm XY}^{\rm bis}(k)
		=\Gamma_{({\rm X}}^{(1)}({\bf k})\int\frac{\mathrm{d}^3{\bf p}}{(2\pi )^3}\Gamma_{{\rm Y})}^{(2)}({\bf p},{\bf k}-{\bf p})B_{\rm L}({\bf k},-{\bf p},-{\bf k}+{\bf p})
	\,,\\
	&&P_{\rm XY}^{\rm tri}(k)
		=\frac{1}{4}\int\frac{\mathrm{d}^3{\bf p}_1}{(2\pi)^3}\int\frac{\mathrm{d}^3{\bf p}_2}{(2\pi)^3}\Gamma_{({\rm X}}^{(2)}({\bf p}_1,{\bf k}-{\bf p}_1)\Gamma_{{\rm Y})}^{(2)}(-{\bf p}_2,-{\bf k}+{\bf p}_2)
	\nonumber\\
	&&\qquad\qquad\qquad\times
			T_{\rm L}({\bf p}_1,{\bf k}-{\bf p}_1,-{\bf p}_2,-{\bf k}+{\bf p}_2)
	\nonumber\\
	&&\quad\quad\quad\quad
		+\frac{1}{3}\Gamma_{({\rm X}}^{(1)}({\bf k})\int\frac{\mathrm{d}^3{\bf p}_1}{(2\pi)^3}\int\frac{\mathrm{d}^3{\bf p}_2}{(2\pi)^3}\Gamma_{{\rm Y})}^{(3)}({\bf p}_1,{\bf p}_2,{\bf k}-{\bf p}_1-{\bf p}_2)
	\nonumber\\
	&&\qquad\qquad\qquad\times
			T_{\rm L}({\bf k},-{\bf p}_1,-{\bf p}_2,-{\bf k}+{\bf p}_1+{\bf p}_2)
	\,,
}
with
$\Gamma^{(n)}_{\rm X}$ 
being the multipoint propagators defined in Eq.~(\ref{eq:multipoint propagator def}).
Since the multipoint propagators are fully non-perturbative quantities and include the non-linear gravitational evolution and halo/galaxy bias properties,
it is difficult to evaluate them over all scale.  However, for the large scale of our interest, these can reduce to simpler expressions.
Taking the large-scale limit, that is $k\ll p$, we find that the leading terms can be described in terms of the renormalized bias functions Eq.~(\ref{eq:c_n def}) as~\citep{Yokoyama:2012az}
\al{
	&&P_{\rm XY}^{\rm tree}(k)
		\approx b_{\rm X}^{(1)}b_{\rm Y}^{(1)}P_{\rm L}(k)
	\,,\\
	&&P_{\rm XY}^{\rm bis}(k)
		\approx b_{({\rm X}}^{(1)}\int\frac{\mathrm{d}^3{\bf p}}{(2\pi )^3}c_{{\rm Y})}^{(2)}({\bf p},-{\bf p})B_{\rm L}({\bf k},-{\bf p},{\bf p})
	\,,\\
	&&P_{\rm XY}^{\rm tri}(k)
		\approx\frac{1}{4}\int\frac{\mathrm{d}^3{\bf p}_1}{(2\pi)^3}\int\frac{\mathrm{d}^3{\bf p}_2}{(2\pi)^3}c_{\rm X}^{(2)}({\bf p}_1,-{\bf p}_1)c_{\rm Y}^{(2)}(-{\bf p}_2,{\bf p}_2)
			T_{\rm L}({\bf p}_1,{\bf k}-{\bf p}_1,-{\bf p}_2,-{\bf k}+{\bf p}_2)
	\nonumber\\
	&&\quad\quad\quad\quad
		+\frac{1}{3}b_{({\rm X}}^{(1)}\int\frac{\mathrm{d}^3{\bf p}_1}{(2\pi)^3}\int\frac{\mathrm{d}^3{\bf p}_2}{(2\pi)^3}c_{{\rm Y})}^{(3)}({\bf p}_1,{\bf p}_2,-{\bf p}_1-{\bf p}_2)
			T_{\rm L}({\bf k},-{\bf p}_1,-{\bf p}_2,{\bf p}_1+{\bf p}_2)
	\,,
}
where $b_{\rm X}^{(1)}\equiv\Gamma_{\rm X}^{(1)}(k\to 0)=1+b_1^{\rm L}$ corresponds to the linear halo bias.
When we consider the generalized local form of the primordial bispectrum and trispectrum,  Eqs.~(\ref{eq:generalized local bis})--(\ref{eq:generalized local tri}), 
the leading order contributions of the power spectrum of the biased objects X and Y 
are given by
\al{
    P_{\rm XY}(k)
        \approx\biggl[ &&
            b_{({\rm X}}^{(1)}b_{{\rm Y})}^{(1)}
            +2b_{({\rm X}}
            \left(f_{\rm NL}\Delta b^f_{{\rm Y})}(k)
            +g_{\rm NL}\Delta b^g_{{\rm Y})}(k)\right)
	\nonumber\\
	&&\qquad
            +\frac{25\tau_{\rm NL}}{36\xi_{\rm m}(k)}
            \Delta b^f_{\rm X}(k)\Delta b^f_{\rm Y}(k)
            \biggr] P_{\rm L}(k)
	\,.
}
Here the bias corrections due to the primordial bi- and tri-spectra
are written as (see also \citep{Yokoyama:2012az})
\al{
	&&\Delta b^f_{\rm X}(k)
		=\frac{2}{{\cal M}(k)}
		\xi_{\rm m}(k)
		\int\frac{\mathrm{d}^3{\bf p}}{(2\pi)^3}c_{\rm X}^{(2)}({\bf p},-{\bf p};M)
		\xi_{\rm m}(p)\xi_{\rm s}(p)P_{\rm L}(p)
	\,,\label{eq:Delta b}\\
	&&\Delta b^g_{\rm X}(k)
	   =\frac{3}{{\cal M}(k)}
	   \xi_{\rm m}(k)
	   \int\frac{\mathrm{d}^3{\bf p}_1}{(2\pi)^3}\int\frac{\mathrm{d}^3{\bf p}_2}{(2\pi)^3}
	   c_{\rm X}^{(3)}({\bf p}_1,{\bf p}_2,-{\bf p}_1-{\bf p}_2)
    \nonumber\\
    &&\quad\quad\quad\quad\quad\times
	   \xi_{\rm m}(p_1)\xi_{\rm m}(p_2)\xi_{\rm g}(|{\bf p}_1+{\bf p}_2|)
	   {\cal M}(p_1){\cal M}(p_2){\cal M}(|{\bf p}_1+{\bf p}_2|)P_\Phi (p_1)P_\Phi (p_2)
	\,.\label{eq:Delta b g}
}
Substituting the definition of the second- and third-order renormalized bias functions Eq.~(\ref{eq:c_n def}) into Eqs.~(\ref{eq:Delta b}) and (\ref{eq:Delta b g}), we have
\al{
	&&\Delta b^f_{\rm X}(k)
	    =\frac{2\beta^{(2)}_{n_{\rm NL}^{\rm s}+n_{\rm NL}^{\rm m}}(M)}{{\cal M}(k)}
	    \left(\frac{k}{k_{\rm piv}}\right)^{n_{\rm NL}^{\rm m}}
	\,,\\
	&&\Delta b^g_{\rm X}(k)
	    =\frac{\beta^{(3)}_{n_{\rm NL}^{\rm m},n_{\rm NL}^{\rm m},n_{\rm NL}^{\rm g}}(M)}{2{\cal M}(k)}
	    \left(\frac{k}{k_{\rm piv}}\right)^{n_{\rm NL}^{\rm m}}
	 \,.
}
Here, the coefficients $\beta^{(2)}_{\alpha}$ and $\beta^{(3)}_{\alpha\beta\gamma}$ are defined as
\al{
	&&\beta^{(2)}_{\alpha}(M)
	    :=\frac{\sigma_\alpha^2(M)}{\delta_{\rm c}^2}
			\biggl[
			    A_2(M)
			    +2A_1(M)\left(\frac{\partial\ln\sigma_\alpha (M)}{\partial\ln\sigma (M)}-1\right)
			\biggr]
	\,,\label{eq:beta_alpha def}\\
	&&\beta^{(3)}_{\alpha\beta\gamma} (M)
	    :=\frac{\sigma^4(M) S_{3,\alpha\beta\gamma}(M)}{\delta_{\rm c}^3}
	        \biggl[
	            A_3(M)
	            +A_2(M)\left(\frac{\mathrm{d}\ln S_{3,\alpha\beta\gamma}(M)}{\mathrm{d}\ln\sigma (M)}+1\right)
	        \biggr]
	 \,,
}
where $\delta_c$ is the critical overdensity, $A_1(M)$, $A_2(M)$, and $A_3(M)$ can be obtained from Eqs.~(\ref{eq:A1})--(\ref{eq:A3}), and we have introduced the modified variance and skewness of the smoothed density field, which are given by
\al{
	&&\sigma_\alpha^2(M)
		=\int\frac{\mathrm{d}^3{\bf k}}{(2\pi )^3}W^2(kR)\left(\frac{k}{k_{\rm piv}}\right)^\alpha P_{\rm L} (k)
	\,,\\
	&&S_{3,\alpha\beta\gamma}(M)
	    =\frac{6}{\sigma^4(M)}\int\frac{\mathrm{d}^3{\bf p}_1}{(2\pi)^3}\int\frac{\mathrm{d}^3{\bf p}_2}{(2\pi)^3}
	        W(p_1R)W(p_2R)W(|{\bf p}_1+{\bf p}_2|R)
	\nonumber\\
	&&\quad\quad\times
	        \left(\frac{p_1}{k_{\rm piv}}\right)^\alpha
	        \left(\frac{p_2}{k_{\rm piv}}\right)^\beta
	        \left(\frac{|{\bf p}_1+{\bf p}_2|}{k_{\rm piv}}\right)^\gamma
	        {\cal M}(p_1){\cal M}(p_2){\cal M}(|{\bf p}_1+{\bf p}_2|)
	        P_\Phi (p_1)P_\Phi (p_2) \,.
}
These results shown here are the generalization of the well-known scaling relations to account for the scale-dependent local-type primordial non-Gaussianities.
If we take $n_{\rm NL}^{\rm m}=n_{\rm NL}^{\rm s}=n_{\rm NL}^{\rm g}=0$, we reproduce the standard formula 
of the scale-dependent bias~\citep{Yokoyama:2012az}.
Using the large-scale behavior of the transfer function ${\cal M}\propto k^2$, we analytically estimate the scaling relation of the bias on large scales:
$\Delta b^f, \Delta b^g\propto k^{-2+n_{\rm NL}^{\rm m}}$.
We found that even in the presence of the nonvanishing $n_{\rm NL}^{\rm s}$ and $n_{\rm NL}^{\rm g}$, 
the scale-dependence of $\Delta b^f$ and $\Delta b^g$ is determined only by $n_{\rm NL}^{\rm m}$.
If $n_{\rm NL}^{\rm m}<2$, the effect of the scale-dependence due to the primordial bi- and tri-spectra
is more prominent on large scale and at high redshift.

\subsection{Bispectrum of biased object in iPT}
\label{sec:Galaxy bispectrum}

We then apply the iPT framework to the bispectrum of biased objects.
The tree-level bispectrum of the biased objects ${\rm X}$, ${\rm Y}$, and ${\rm Z}$ up to the leading order of the nonlinearity parameters can be represented as~\citep{Yokoyama:2013mta}
\al{
	B_{\rm XYZ}({\bf k}_1,{\bf k}_2,{\bf k}_3)
		=B_{\rm XYZ}^{\rm grav}({\bf k}_1,{\bf k}_2,{\bf k}_3)
			+B_{\rm XYZ}^{\rm bis}({\bf k}_1,{\bf k}_2,{\bf k}_3)
			+B_{\rm XYZ}^{\rm tris}({\bf k}_1,{\bf k}_2,{\bf k}_3)
			+\cdots
	\,,
}
where
\al{
	&&B_{\rm XYZ}^{\rm grav}({\bf k}_1,{\bf k_2},{\bf k}_3)
		=\Gamma_{({\rm X}}^{(1)}({\bf k}_1)\Gamma_{\rm Y}^{(1)}({\bf k}_2)\Gamma_{{\rm Z})}^{(2)}(-{\bf k}_1,-{\bf k}_2) P_{\rm L}(k_1)P_{\rm L}(k_2) +(2\ {\rm perms.})
	\,,\\
	&&B_{\rm XYZ}^{\rm bis}({\bf k}_1,{\bf k_2},{\bf k}_3)
		=\Gamma_{\rm X}^{(1)}({\bf k}_1)\Gamma_{\rm Y}^{(1)}({\bf k}_2)\Gamma_{\rm Z}^{(1)}({\bf k}_3)B_{\rm L}(k_1,k_2,k_3)
	\,,\\
	&&B_{\rm XYZ}^{\rm tri}({\bf k}_1,{\bf k_2},{\bf k}_3)
		=\frac{1}{2}\Gamma_{({\rm X}}^{(1)}({\bf k}_1)\Gamma_{\rm Y}^{(1)}({\bf k}_2)\int\frac{\mathrm{d}^3{\bf p}}{(2\pi)^3}\Gamma_{{\rm Z})}^{(2)}({\bf p},{\bf k}_3-{\bf p})T_{\rm L}({\bf k}_1,{\bf k}_2,{\bf p},{\bf k}_3-{\bf p})
	\nonumber\\
	&&\qquad\qquad\qquad\qquad
			+(2\ {\rm perms.})
	\,.\label{eq:gal bi tri}
}
Following the same step as the previous subsection, we take the large-scale limit to reduce the above expression.
The leading terms are given by
\al{
	&&B_{\rm XYZ}^{\rm grav}({\bf k}_1,{\bf k_2},{\bf k}_3)
		\approx b_{({\rm X}}^{(1)}b_{\rm Y}^{(1)}\widetilde\Gamma_{{\rm Z})}^{(2)}(-{\bf k}_1,-{\bf k}_2) P_{\rm L}(k_1)P_{\rm L}(k_2) +(2\ {\rm perms.})
	\,,\\
	&&B_{\rm XYZ}^{\rm bis}({\bf k}_1,{\bf k_2},{\bf k}_3)
		\approx b_{\rm X}^{(1)}b_{\rm Y}^{(1)}b_{\rm Z}^{(1)}B_{\rm L}(k_1,k_2,k_3)
	\,,\\
	&&B_{\rm XYZ}^{\rm tri}({\bf k}_1,{\bf k_2},{\bf k}_3)
		\approx\frac{1}{2}b_{({\rm X}}^{(1)}b_{\rm Y}^{(1)}\int\frac{\mathrm{d}^3{\bf p}}{(2\pi)^3}c_{{\rm Z})}^{(2)}({\bf p},-{\bf p})T_{\rm L}({\bf k}_1,{\bf k}_2,{\bf p},{\bf k}_3-{\bf p})
			+(2\ {\rm perms.})
	\,,\label{eq:biased bi tri}
}
where $\widetilde\Gamma^{(2)}_{\rm X}$ is the large-scale limit of the second-order multipoint propagator:
\al{
	\widetilde\Gamma^{(2)}_{\rm X}({\bf k}_1,{\bf k}_2)
		=F_2({\bf k}_1,{\bf k}_2)+\biggl[ 2+\left(\frac{k_1}{k_2}+\frac{k_2}{k_1}\right)\frac{{\bf k}_1\cdot{\bf k}_2}{k_1k_2}\biggr] b_1^{\rm L} +b_2^{\rm L}
	\,.
}
With the generalized local form of the primordial bispectrum Eq.~(\ref{eq:generalized local bis}), we then write down the biased bispectum induced by the primordial one
\al{
	B_{\rm XYZ}^{\rm bis}
		\approx f_{\rm NL}b_{\rm X}^{(1)}b_{\rm Y}^{(1)}b_{\rm Z}^{(1)}\widehat B_{\rm L}(k_1,k_2,k_3)+(2\ {\rm perms.})
	\,.\label{eq:B_bis}
}
Here we have introduced the linear bispectrum with $f_{\rm NL}=1$, which is defined as
\al{
	\widehat B_{\rm L}(k_1,k_2,k_3)
		=&&2{\cal M}(k_1){\cal M}(k_2){\cal M}(k_3)\xi_{\rm m}(k_1)\xi_{\rm m}(k_2)\xi_{\rm s}(k_3)P_\Phi (k_1)P_\Phi (k_2)
	\nonumber\\
		=&&2 \frac{{\cal M}(k_3)\xi_{\rm m}(k_1)\xi_{\rm m}(k_2)\xi_{\rm s}(k_3)}{{\cal M}(k_1) {\cal M} (k_2)}P_{\rm L} (k_1)P_{\rm L} (k_2)
	\,.
}
Substituting the generalized local form Eq.~(\ref{eq:generalized local tauNL})
into Eq.~(\ref{eq:biased bi tri}), we obtain the biased bispectrum induced
by the $\tau_{\rm NL}$ term as
\al{
    B^{\tau_{\rm NL}}_{\rm XYZ}
    =&&b_{\rm X}^{(1)}b_{\rm Y}^{(1)}
        \frac{25\tau_{\rm NL}}{18{\cal M}(k_3)}
	\nonumber\\
	&&\times
        \biggl[
            \frac{\beta_{2n_{\rm NL}^{\rm s}+n_{\rm NL}^{\rm m}}}{\xi_{\rm s}(k_3)}\widehat B_{\rm L}(k_1,k_2,k_3)
            +\beta_{n_{\rm NL}^{\rm s}+n_{\rm NL}^{\rm m}}
           \left(\widehat B_{\rm L}(k_2,k_3,k_1)+\widehat B_{\rm L}(k_3,k_1,k_2)\right)
        \biggr]
	\nonumber\\
	&& +(2\ {\rm perms.})
    \nonumber\\
    =&&b_{({\rm X}}^{(1)}b_{\rm Y}^{(1)}\Delta\widetilde b^\tau_{Z)}(k_1,k_2,k_3)
        \widehat B_{\rm L}(k_1,k_2,k_3)
        +(2\ {\rm perms.})
    \,,
}
where the bias correction is expressed as
\al{
    \Delta\widetilde b^\tau_{\rm X}(k_1,k_2,k_3)
        =\frac{25}{18}\tau_{\rm NL}
            \biggl[
                \beta_{n_{\rm NL}^{\rm s}+n_{\rm NL}^{\rm m}}
                    \left(\frac{1}{{\cal M}(k_1)}+\frac{1}{{\cal M}(k_2)}\right)
                    +\frac{\beta_{2n_{\rm NL}^{\rm s}+n_{\rm NL}^{\rm m}}}{\xi_{\rm s}(k_3){\cal M}(k_3)}
            \biggr]
    \,.
}
Here, $\beta_\alpha$ was defined in Eq.~(\ref{eq:beta_alpha def}).
The bispectrum induced by the $g_{\rm NL}$ term is given by
\al{
    B^{g_{\rm NL}}_{\rm XYZ}
        =b_{({\rm X}}^{(1)}b_{\rm Y}^{(1)}\Delta\widetilde b^g_{{\rm Z})}(k_3)
        \widehat B_{\rm L}(k_1,k_2,k_3)
        +(2\ {\rm perms.})
    \,,
}
with
\al{
    \Delta\widetilde b^g_{\rm X}(k_3)=3g_{\rm NL}\frac{\beta_{n_{\rm NL}^{\rm g}+n_{\rm NL}^{\rm m}}}{\xi_{\rm s}(k_3){\cal M}(k_3)}
    \,.
}
We found that the contributions from the primordial trispectrum provide
the additional scale-dependence 
compared to that from the primordial bispectrum.
We should emphasize that the dependence of the bias on $n_{\rm NL}^{\rm A}$ in the biased bispectrum 
is different from that of the bias for the power spectrum.
In particular, there appears the explicit scale-dependence with the power-law index $n_{\rm NL}^{\rm s}$, unlike
the biased power spectrum. 
Hence, we expect that this can break the degeneracy between the scale-dependence of 
the primordial non-Gaussianities.

\section{Fisher analysis}
\label{sec:Results}

\subsection{Observable of galaxy clustering and error covariance}

To connect the halo formalism with galaxy clustering, we need a method for describing the way how galaxies populate the dark matter halos.
In this paper, we adopt a halo model~\citep{Cooray:2002dia}, in which we use 
the Halo Occupation Distribution (HOD) functions to provide the mean number density of galaxies
per a halo of a given mass, $\ave{N(M,z)}$.
We estimate the redshift distribution of number density of galaxies per unit area, $n_{\rm g}(z)$,
through the weighted average of the halo mass function $n(M,z)$ as
\al{
	n_{\rm g}(z)=\frac{\mathrm{d}^2 V}{\mathrm{d} z\mathrm{d}\Omega}\int_{M_{\rm min}}^\infty\mathrm{d} M\, n(M,z)\ave{N(M,z)}
	\,,
}
where $\mathrm{d}^2 V/\mathrm{d}\Omega\mathrm{d} z=\chi^2(z)/H(z)$ denotes the comoving volume element per unit redshift per unit steradian, 
with $\chi$ denoting the comoving radial distance. 
We then define the projected number density of galaxies in the redshift interval $\overline z_i-\Delta z/2<z<\overline z_i+\Delta z/2$ as
\al{
	\overline n_{\rm g}(\overline z_i)=\int_{\overline z_i-\Delta z/2}^{\overline z_i+\Delta z/2}\mathrm{d} z\,n_{\rm g}(z)
	\,.
} 
Using this, we set $M_{\rm min}$ so that $\overline n_{\rm g}(\overline z_i)$ matches the expected mean galaxy number density of the survey in each redshift bin.
As for the HOD function, we adopt the model proposed by \citep{Tinker:2004gf}, which is given as
\al{
	\ave{N(M,z)}=1+\frac{M}{M_1(z)}\exp\left( -\frac{M_{\rm cut}(z)}{M}\right)
	\,,
}
for $M>M_{\rm min}(z)$ and zero otherwise.
We employ a fit to $N$-body simulations for the mass scale $M_1$ and $M_{\rm cut}$ \citep{Conroy:2005aq}.
In the subsequent analysis, we assume the following correlations between $M_1$\,, $M_{\rm cut}$\,,
and $M_{\rm min}$ : 
$\log_{10}(M_{\rm cut}/h^{-1}M_{\odot})=0.76\log_{10}(M_1/h^{-1}M_\odot)+2.3$
and $\log_{10}(M_1/M_{\rm min})=1.1$.
\\

In our analysis, we adopt the angular power- and bi-spectra for galaxy clustering as relevant statistical quantities.
Employing the Limber approximation, the angular power- and bi-spectra 
in the $i$-th redshift bin are respectively expressed as
\al{
	C_i(\ell )=&&\int_{\overline z_i-\Delta z/2}^{\overline z_i+\Delta z/2}\mathrm{d} z\frac{H(z)}{\chi^2(z)}
	\nonumber\\
	&&\times\prod_{{\rm m}=1,2}
			\biggl[\frac{1}{\overline n_{\rm g}(\overline z_i)}\frac{\mathrm{d}^2 V}{\mathrm{d} z\mathrm{d}\Omega}\int_{M_{\rm min}(\overline z_i)}^\infty\mathrm{d} M_mn(M_m,z)\ave{N(M_m,z)}\biggr]
			P_{\rm gg}\left(\frac{\ell +1/2}{\chi(z)};z\right)
	\,,
}
and
\al{
	B_i(\ell_1,\ell_2,\ell_3)
		=&&\int_{\overline z_i-\Delta z/2}^{\overline z_i+\Delta z/2}\mathrm{d} z\frac{H^2(z)}{\chi^4(z)}
	\nonumber\\
	&&\times
		\prod_{{\rm m}=1,2,3}
			\biggl[\frac{1}{\overline n_{\rm g}(\overline z_i)}\frac{\mathrm{d}^2 V}{\mathrm{d} z\mathrm{d}\Omega}\int_{M_{\rm min}(\overline z_i)}^\infty\mathrm{d} M_mn(M_m,z)\ave{N(M_m,z)}\biggr]
	\nonumber\\
	&&
		\times B_{\rm ggg}\left(\frac{\ell_1 +1/2}{\chi(z)},\frac{\ell_2+1/2}{\chi(z)},\frac{\ell_3+1/2}{\chi(z)};z\right)
	\,.
}
The three-dimensional power- and bi-spectra of galaxy clustering, $P_{\rm gg}$ and $B_{\rm ggg}$, are obtained from Secs.~\ref{sec:Galaxy power spectrum} and \ref{sec:Galaxy bispectrum}.
Using the resultant formulae, we adopt the Fisher analysis to estimate expected errors of model parameters for a given survey.
For simplicity, we will neglect the cross covariance between the power
and bi-spectra. 
The Fisher matrix is defined as
\al{
	F_{\alpha\beta}=F_{\alpha\beta}^P+F_{\alpha\beta}^B
	\,,
}
where
\al{
	&&F_{\alpha\beta}^P=\sum_{\ell,\ell^\prime =\ell_{\rm min}}^{\ell_{\rm max}}\sum_{i,j}
				\frac{\partial C_i(\ell )}{\partial\theta^\alpha}{\rm Cov}\Bigl[{\bf C}(\ell),{\bf C}(\ell^\prime )]\Bigr]^{-1}_{ij}\frac{\partial C_j(\ell^\prime )}{\partial\theta^\beta}
	\,,\\
	&&F_{\alpha\beta}^B=\sum_{\ell,\ell^\prime =\ell_{\rm min}}^{\ell_{\rm max}}\sum_{i,j}
				\frac{\partial B_i(\ell_1,\ell_2,\ell_3)}{\partial\theta^\alpha}{\rm Cov}\Bigl[{\bf B}(\ell_1,\ell_2,\ell_3),{\bf B}(\ell_1^\prime,\ell_2^\prime.\ell_3^\prime )]\Bigr]^{-1}_{ij}\frac{\partial B_j(\ell_1^\prime,\ell_2^\prime,\ell_3^\prime )}{\partial\theta^\beta}
	\,.
}
Assuming the Gaussian error covariance, the covariant matrices of the angular power and bi-spectra are expressed as~\citep{Kayo:2012nm}
\al{
	&&{\rm Cov}[C_{\ell}^i,C_{\ell^\prime}^j]
		=2\frac{\delta_{ij}^{\rm K}\delta^{\rm K}_{\ell\ell^\prime}}{N_{\rm pair}(\ell )}\left( C_i(\ell )+\frac{1}{\overline n_{\rm g}(\overline z_i)}\right)^2
	\,,\\
	&&{\rm Cov}[B_i(\ell_1,\ell_2,\ell_3),B_j(\ell_1^\prime,\ell_2^\prime,\ell_3^\prime )]
	\nonumber\\
	&&\qquad
		=\frac{\Omega_{\rm s}\delta^{\rm K}_{ij}}{N_{\rm trip}(\ell_1,\ell_2,\ell_3)}\left( C_i(\ell )+\frac{1}{\overline n_{\rm g}(\overline z_i)}\right)^3
				\biggl[\delta^{\rm K}_{\ell_1\ell_1^\prime}\delta^{\rm K}_{\ell_2\ell_2^\prime}\delta^{\rm K}_{\ell_3\ell_3^\prime}+(5\,{\rm perms})\biggr]
	\,.
}
Here the quantity $N_{\rm pair}(\ell)$ denotes the number of independent pairs for two vectors ${\bf\ell}$ and $-{\bf\ell}$ within
the multipole bin $[\ell -\Delta\ell/2,\ell +\Delta\ell /2]$, and
$N_{\rm trip}(\ell_1,\ell_2,\ell_3)$ is the number of independent triplets for three vectors forming triangular configuration.
The explicit expressions of these quantities are given by
\al{
	&&N_{\rm pair}(\ell )\approx\frac{2\pi\ell\Delta\ell}{\ell_{\rm f}^2}
	\,,\\
	&&N_{\rm trip}(\ell_1,\ell_2,\ell_3)
		\approx\frac{1}{\ell_{\rm f}^4}\frac{8\pi\ell_1\ell_2\ell_3\Delta\ell_1\Delta\ell_2\Delta\ell_3}{\sqrt{2\ell_1^2\ell_2+2\ell_2\ell_3^2+2\ell_3^2\ell_1^2-\ell_1^4-\ell_2^4-\ell_3^4}}
	\,,
}
where $\ell_{\rm f}=2\pi/\sqrt{\Omega_{\rm s}}$ is the fundamental multipole with $\Omega_{\rm s}=4\pi f_{\rm sky}$ being the survey area in unit of steradian. 
In the subsequent analysis, we take $\Delta\ell =\Delta\ell_1=\Delta\ell_2=\Delta\ell_3=\ell_{\rm min}=\ell_{\rm f}$,
while the maximum multipole $\ell_{\rm max}$ is set to $150$. We note that this may lead to the conservative constraints, since
for galaxy bispectrum, increasing $\ell_{\rm max}$ would potentially give tighter constraints. 
However, as increasing $\ell_{\rm max}$ the small-scale contributions to the galaxy bispectrum becomes rather important,
hence we take the conservative choice.
Here, we consider $6$ parameters in the Fisher analysis; $3$ parameters for the primordial bispectrum, $(f_{\rm NL},n_{\rm NL}^{\rm s},n_{\rm NL}^{\rm m})$, 3 parameters for the primordial trispectrum, 
$(\tau_{\rm NL}, g_{\rm NL},n_{\rm NL}^{\rm g})$ with the fiducial values 
$(f_{\rm NL},n_{\rm NL}^{\rm s},n_{\rm NL}^{\rm m},\tau_{\rm NL}, g_{\rm NL},n_{\rm NL}^{\rm g})=(1,0,0,36/25,1,0)$.
Furthermore, we do not marginalize the uncertainty in the halo/galaxy bias properties.
Since the clustering bias is expected to be observationally determined at the relatively small scales, in which
the effect of the primordial non-Gaussianities is not so large, we expect that there is no serious parameter degeneracy with the nonlinearity parameters.

\subsection{Results}

Based on the derived analysis tool, we then apply our Fisher matrix analysis to future galaxy surveys as a demonstration.
To elucidate the statistical power of power and bi-spectra, we shall consider two representative galaxy surveys: SKAO and Euclid.
We adopt the predicted number density of galaxies as a function of redshift, given in Appendix B of
Ref.~\citep{Bull:2015lja} for SKAO and Table 3 of Ref.~\citep{Amendola:2016saw} for Euclid,
respectively.

\begin{figure}
\begin{center}
\vspace{-35mm}
\includegraphics[width=175mm]{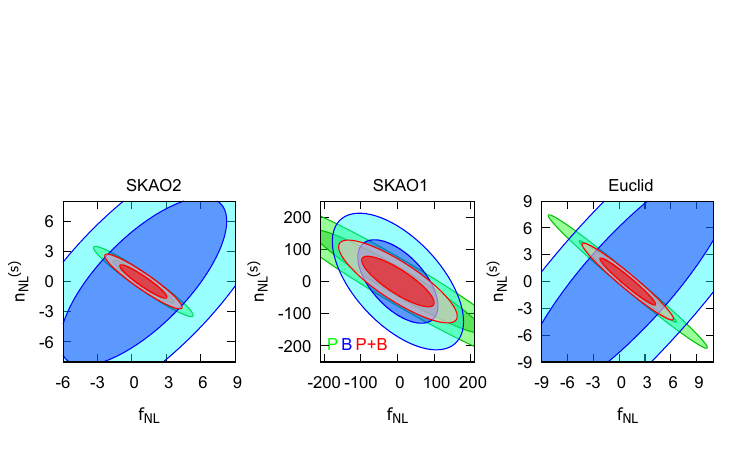}
\vspace{-10mm}
\end{center}
\caption{
The marginalized $1$ and $2\sigma$ error contours on $(f_{\rm NL},n_{\rm NL}^{\rm s})$ in the single-source model for SKAO2 (left), SKAO1 (middle) and Euclid (right). For each panel, the cases with power spectrum alone (green), bispectrum alone (blue), combined analyses (red) are shown.
}
\label{fig:plot_fNL-ns}
\end{figure} 

Let us consider the single-source model, which was discussed in Sec.~\ref{sec:Generic single-source case}.
Since we can take $n_{\rm NL}^{\rm m}=0$ and $\tau_{\rm NL}$ can be rewritten 
in terms of $f_{\rm NL}$ and $n_{\rm NL}^{\rm s}$, 
the number of independent parameters
can reduce to four, 
$(f_{\rm NL},n_{\rm NL}^{\rm s}, g_{\rm NL}, n_{\rm NL}^{\rm g})$.
In order to simplify the analysis, we further assume that only 
$f_{\rm NL}$ and $n_{\rm NL}^{\rm s}$ are taken into account.
The two-dimensional error contours on $(f_{\rm NL},n_{\rm NL}^{\rm s})$ are shown in Fig.~\ref{fig:plot_fNL-ns},
and the marginalized (unmarginalized) $1\sigma$ errors on them are summarized in Table \ref{table:single-source model}.
These figure and table imply that the expected
constraints 
from the galaxy power spectrum alone provide a severe
degeneracy 
between $f_{\rm NL}$ and $n_{\rm NL}^{\rm s}$, hence the marginalized constraint are 
substantially degraded, compared to the unmarginalized ones. 
This can be understood as follows: As shown in Sec.~\ref{sec:Galaxy power spectrum},
$n_{\rm NL}^{\rm s}$ does not give any scale-dependence to the bias factor of the galaxy power spectrum, 
but there appears a weak dependence of $n_{\rm NL}^{\rm s}$ 
on the effective variation of the smoothed density fluctuations.
On the other hand, as for the large-scale galaxy bispectrum, 
although its constraining power is relatively weak, 
there is a clear scale-dependence with the power-law index $n_{\rm NL}^{\rm s}$.
Therefore, we found that it would be useful to use the information of the galaxy bispectrum 
to break the degeneracy between them and obtain the tighter constraints 
on both $f_{\rm NL}$ and $n_{\rm NL}^{\rm s}$.

\begin{table}[tb]
\begin{tabular}{|c|c|c|c|} \hline
SKAO2 & $P$ alone  & $B$ alone & $P+B$ combined  \\ \hline
$\sigma (f_{\rm NL})$ & $1.7(0.48)$ & $4.8(3.3)$ & $1.4(0.47)$ \\
$\sigma (n_{\rm NL}^{\rm s})$ & $1.4(0.38)$ & $6.0(3.8)$ & $1.1(0.38)$ \\
\hline\hline
SKAO1 & $P$ alone  & $B$ alone & $P+B$ combined  \\ \hline
$\sigma (f_{\rm NL})$ & $1.5\times 10^2(45)$ & $1.0\times 10^2 (58)$ & $85(36)$ \\
$\sigma (n_{\rm NL}^{\rm s})$ & $1.1\times 10^2(31)$ & $1.1\times 10^2(70)$ & $64(28)$ \\
\hline\hline
Euclid & $P$ alone  & $B$ alone & $P+B$ combined  \\ \hline
$\sigma (f_{\rm NL})$ & $3.7(0.54)$ & $8.6(4.4)$ & $2.2(0.53)$ \\
$\sigma (n_{\rm NL}^{\rm s})$ & $3.0(0.43)$ & $9.8(4.7)$ & $1.8(0.43)$ \\
\hline
\end{tabular}
\vspace{3mm}
\caption{Forecast results of marginalized (unmarginalized) $1\sigma$ errors on primordial 
non-Gaussianity parameters for the case of the single-source model, 
which is discussed in Sec.~\ref{sec:Generic single-source case}.
}
\label{table:single-source model}
\end{table}

Next, we consider the model of the generic scale-dependent primordial non-Gaussianities, 
in which six independent parameters are considered.
In Table \ref{table:generic model}, we summarize the marginalized (unmarginalized) errors
on $(f_{\rm NL},n_{\rm NL}^{\rm s},n_{\rm NL}^{\rm m},\tau_{\rm NL}, g_{\rm NL},n_{\rm NL}^{\rm g})$.
There still exists the degeneracy among these parameters
for the case of the galaxy power spectrum in some complicated way.
Unfortunately, the parameter degeneracy also appears for the galaxy bispectrum, and the constraining power becomes worse.
However, the combined analysis between the galaxy power spectrum and bispectrum can be
used to break the parameter degeneracy and obtain the tighter constraints.

\begin{table}[tb]
\begin{tabular}{|c|c|c|c|} \hline
SKAO2 & $P$ alone  & $B$ alone & $P+B$ combined  \\ \hline
$\sigma (f_{\rm NL})$ & $11(0.53)$ & $18(3.4)$ & $3.5(0.53)$ \\
$\sigma (n_{\rm NL}^{\rm s})$ & $7.9(0.38)$ & $10(3.8)$ & $1.5(0.38)$ \\
$\sigma (n_{\rm NL}^{\rm m})$ & $1.3(0.20)$ & $7.1(0.84)$ & $1.1(0.19)$ \\
$\sigma (\tau_{\rm NL})$ & $38(11)$ & $5.7\times 10^2(2.1\times 10^2)$ & $33(11)$ \\
$\sigma (g_{\rm NL} )$ & $1.3\times 10^6(5.7\times 10^4)$ & $9.5\times 10^3(2.7\times 10^3)$ & $8.7\times 10^3(2.7\times 10^3)$ \\
$\sigma (n_{\rm NL}^{\rm g})$ & $1.9\times 10^5(8.0\times 10^3)$ & $6.4\times 10^3(1.9\times 10^3)$ & $5.8\times 10^3(1.8\times 10^3)$ \\ \hline\hline
SKAO1 & $P$ alone  & $B$ alone & $P+B$ combined  \\ \hline
$\sigma (f_{\rm NL})$ & $9.5\times 10^2(45)$ & $1.9\times 10^2(59)$ & $1.2\times 10^2(36)$ \\
$\sigma (n_{\rm NL}^{\rm s})$ & $9.4\times 10^2(31)$ & $1.9\times 10^3(69)$ & $1.1\times 10^2(28)$ \\
$n_{\rm NL}^{\rm m}$ & $88(35)$ & $1.4\times 10^2(35)$ & $71(25)$ \\
$\sigma (\tau_{\rm NL})$ & $1.5\times 10^5(4.8\times 10^4)$ & $2.0\times 10^5(5.7\times 10^4)$ & $1.1\times 10^5(3.7\times 10^4)$ \\
$\sigma (g_{\rm NL})$ & $3.5\times 10^7(1.2\times 10^6)$ & $9.6\times 10^5(2.3\times 10^5)$ & $6.8\times 10^5(2.3\times 10^5)$ \\
$\sigma (n_{\rm NL}^{\rm g})$ & $7.0\times 10^6(1.8\times 10^5)$ & $6.5\times 10^5(1.6\times 10^5)$ & $3.9\times 10^5(1.2\times 10^5)$ \\ \hline\hline
Euclid & $P$ alone  & $B$ alone & $P+B$ combined  \\ \hline
$\sigma (f_{\rm NL})$ & $27(0.61)$ & $29(4.6)$ & $4.5(0.61)$ \\
$\sigma (n_{\rm NL}^{\rm s})$ & $11(0.43)$ & $15(4.7)$ & $2.2(0.43)$ \\
$\sigma (n_{\rm NL}^{\rm m})$ & $1.6(0.22)$ & $11(1.0)$ & $1.4(0.22)$ \\
$\sigma (\tau_{\rm NL})$ & $47(11)$ & $7.6\times 10^2(2.4\times 10^2)$ & $36(10)$\\
$\sigma (g_{\rm NL})$ & $6.8\times 10^6(1.5\times 10^5)$ & $2.1\times 10^4(3.1\times 10^3)$ & $2.1\times 10^4(3.1\times 10^3)$ \\
$\sigma (n_{\rm NL}^{\rm g})$ & $5.4\times 10^5(2.3\times 10^4)$ & $1.6\times 10^4(2.4\times 10^3)$ & $1.6\times 10^4(2.4\times 10^3)$\\ \hline
\end{tabular}
\vspace{3mm}
\caption{Forecast results of marginalized (unmarginalized) $1\sigma$ errors on primordial non-Gaussianity parameters for the case of the generic model.
}
\label{table:generic model}
\end{table}

\section{Summary}
\label{sec:Summary}

We have revisited the local-type primordial non-Gaussianities and have developed
the formalism which allows us to include their scale-dependence due to
the evolution of the scalar field perturbations after horizon exit and the different scale-dependence of
various fields, based on the $\delta N$ formalism. 
We then have derived the generalized form of the local type
bi- and tri-spectra
[Eqs.~(\ref{eq:fNL def})--(\ref{eq:Pi def}), (\ref{eq:tauNL sol})--(\ref{eq:Theta_IJK def})].
Although the results are generic formulae, these are a bit complicated and may not be suitable 
to compare with observational data.
Therefore, we have proposed  
simpler but phenomenologically meaningful expressions, 
Eqs.~(\ref{eq:generalized local bis})--(\ref{eq:generalized local tri}), which capture
a wide range of physically motivated models.
We then have applied our formulae to the specific models,
and
explicitly shown that not only the amplitude but also 
the scale-dependence 
of the primordial non-Gaussianities can be well parametrized by using 
the generalized local-form.

To see the clustering features of density field caused by the resultant scale-dependent primordial non-Gaussianities, 
adopting the integrated perturbation theory which systematically includes the late-time gravitational
evolution of the effect of halo/galaxy bias, we have derived the formulae of 
the large-scale power spectrum and bispectrum of the biased objects.
We have shown that 
the scale-dependence of the bias factor for the biased power spectrum
is determined only by the power-law index characterizing the multi-field model, $n_{\rm NL}^{\rm m}$, hence
the biased power spectrum analysis cannot accurately constrain
the power-law indexes characterizing the single-source model, $n_{\rm NL}^{\rm s}$
and $n_{\rm NL}^{\rm g}$.
In contrast, the biased bispectrum generated
from the primordial one 
explicitly depends 
on $n_{\rm NL}^{\rm s}$ and $n_{\rm NL}^{\rm g}$, 
therefore the combined analysis with the biased power and bi-spectra
provides the unique opportunity to probe the scale-dependence of the primordial non-Gaussianities.

Performing the Fisher analysis based on the derived formulae, 
we have given the forecasts 
for constraints from the two representative surveys such as SKAO and Euclid.
In the case of the single-source model, the measurements of galaxy power spectrum alone 
provides a severe
degeneracy between $f_{\rm NL}$ and $n_{\rm NL}^{\rm s}$, and
the marginalized constraint is substantially degraded, compared to the unmarginalized ones.
When the galaxy bispectrum is taken into account, there is no serious parameter degeneracy between them
and the marginalized constraint on $n_{\rm NL}^{\rm s}$ approaches to the unmarginalized one.

In this paper, we have made several simplified assumptions.
In particular, we have considered only the large-scale asymptotes of the biased power spectrum
and bispectrum.
In this sense, our results presented in this paper may be regarded as a conservative estimate, and a possibility
to further improve the constraints needs to be investigated for a quantitative estimation.
On the other hand, we have adopted
the Limber-approximation for 
angular power spectrum and bispectrum as 
statistical quantities.
The Limber approximation becomes invalid at lower multipoles, and our treatment with flat-sky limit may not be adequate.
This may be a rather optimistic assumption, however the primary purpose of this paper is to explore the impact of the scale-dependence of the local-type primordial non-Gaussianities 
on future large-scale structure observations.
Hence the full-sky correction at lower multipoles to the galaxy power and bi-spectra is left as a future issue.


\begin{ack}

This work was supported in part by JSPS KAKENHI Grant Nos.~17K14304 (D.Y.), 19H01891 (D.Y.), JP20H01932 (S.Y.), JP20K03968 (S.Y.), 17H01131 (T.T.) and  19K03874 (T.T.), and MEXT KAKENHI Grant No.~19H05110 (T.T.). 

\end{ack}

\appendix

\section{Explicit form of coefficients}
\label{sec:Explicit form of coefficients}

The explicit expressions of the coefficients 
of the super-horizon scalar field perturbations 
in Eqs.~(\ref{eq:delta phi 1})--(\ref{eq:delta phi 3}) 
are given by \citep{Byrnes:2010ft}
\al{
    &&\epsilon_{IJ}=2\sqrt{\epsilon_I\epsilon_J}-\eta_{IJ}
	\label{eq:epsilon_IJ} \,,\\
	&&F^I_{(2)JK}=-\frac{V_{IJK}}{3H^2}-4\sqrt{2\epsilon_I\epsilon_J\epsilon_K}+\eta_{IJ}\sqrt{2\epsilon_K}+\eta_{JK}\sqrt{2\epsilon_I}+\eta_{IK}\sqrt{2\epsilon_J} \label{eq:F(2)}
	\,,\\
	&&F^I_{(3)JKL}=-\frac{V_{IJKL}}{3H^2}
				+\sqrt{2\epsilon_L}\frac{V_{IJK}}{3H^2}+\sqrt{2\epsilon_K}\frac{V_{IJL}}{3H^2}
				+\sqrt{2\epsilon_J}\frac{V_{IKL}}{3H^2}+\sqrt{2\epsilon_I}\frac{V_{JKL}}{3H^2}
	\nonumber\\
	&&\quad\quad\quad\quad\quad\quad
				+24\sqrt{\epsilon_I\epsilon_J\epsilon_K\epsilon_L}-4\left(\eta_{IK}\sqrt{\epsilon_J\epsilon_K}+\eta_{JL}\sqrt{\epsilon_I\epsilon_K}+\eta_{KL}\sqrt{\epsilon_I\epsilon_J}\right)
	\nonumber\\
	&&\quad\quad\quad\quad\quad\quad
				+\eta_{IJ}\eta_{KL}+\eta_{JK}\eta_{IK}+\eta_{IK}\eta_{JL}  \label{eq:F(3)}
	\,.
}
Here we have introduced the slow-roll parameters as $\epsilon =-\dot H/H^2$, $\epsilon_I=(V_I/3H^2)^2/2$, and $\eta_{IJ}=V_{IJ}/3H^2$.

\section{Integrated perturbation theory}
\label{sec:Integrated perturbation theory}

In the iPT framework developed in \citep{Matsubara:2011ck,Matsubara:2012nc}, the power spectrum and bispectrum of the biased tracer are constructed
with multipoint propagators of the biased object ${\rm X}$ in the Eulerian space, $\Gamma_{\rm X}^{(n)}$, which are defined by the ensemble averages 
of the functional derivatives
\al{
	\ave{\frac{\delta^n\delta_{\rm X}({\bf k})}{\delta\delta_{\rm L}({\bf k}_1)\cdots\delta\delta_{\rm L}({\bf k}_n)}}
		=(2\pi)^{3(1-n)}\delta_{\rm D}^3({\bf k}_1+\cdots +{\bf k}_n-{\bf k})\Gamma_{\rm X}^{(n)}({\bf k}_1,\cdots ,{\bf k}_n)
	\,,\label{eq:multipoint propagator def}
}
where $\delta_{\rm L}$ and $\delta_{\rm X}$ denote the linear density field defined in Eq.~(\ref{eq:linear density field}) and 
the number density field of the biased objects in Eulerian space, respectively,
The first- and second-order multipoint propagators can be described as
\al{
	&&\Gamma_{\rm X}^{(1)}({\bf k})\approx 1+c_1^{\rm L}({\bf k})
	\,,\\
	&&\Gamma_{\rm X}^{(2)}({\bf k}_1,{\bf k}_2)\approx F_2({\bf k}_1,{\bf k}_2)+\left( 1+\frac{{\bf k}_1\cdot{\bf k}_2}{k_1^2}\right) c_1^{\rm L}({\bf k}_1) +\left( 1+\frac{{\bf k}_1\cdot{\bf k}_2}{k_2^2}\right) c_1^{\rm L}({\bf k}_2) +c_2^{\rm L}({\bf k}_1,{\bf k}_2)
	\,,
}
where $F_2$ and $c_n^{\rm L}$ are the second-order kernel of the standard perturbation theory and the renormalized bias functions in the Lagragian space.
To proceed the analysis, we need the explicit expression of the renormalized bias functions. 
We then adopt the halo bias prescription proposed in \citep{Matsubara:2012nc}.
In this prescription, the renormalized bias function for halos with mass $M$ can be obtained by the universal mass function, $f_{\rm MF}(\nu )$ with $\nu =\delta_{\rm c}/\sigma (M)$ through
\al{
	c_n^{\rm L}({\bf k}_1,\cdots,{\bf k}_n)
		=\frac{1}{\nu^n}\left( A_n(M)+A_{n-1}(M)\frac{\mathrm{d}}{\mathrm{d}\ln\sigma}\right)\left(\frac{W(k_1R)\cdots W(k_nR)}{\sigma^n}\right)
	\,.\label{eq:c_n def}
}
Here a function $A_n(M)$ is defined as
\al{
	A_n(M)=\sum_{j=0}^n\frac{n!}{j!}\frac{(-\nu )^j}{f_{\rm MF}(\nu)}\frac{\mathrm{d}^j f_{\rm MF}(\nu)}{\mathrm{d}\nu^j}
	\,.\label{eq:A_n def}
}
The halo mass function $n_{\rm h}(M)$ and the scale-independent Lagrangian bias function $b_n^{\rm L}(M)$ can be defined in terms of the universal mass function as
\al{
	&n(M)=-\frac{\rho_{{\rm m},0}}{M^2}\frac{\mathrm{d}\ln\sigma}{\mathrm{d}\ln M}f_{\rm MF}(\nu )
	\,,\label{eq:halo mass function}\\
	&b_n^{\rm L}(M)=\frac{1}{(-\sigma )^nf_{\rm MF}}\frac{\mathrm{d}^n}{\mathrm{d}\nu^n}f_{\rm MF}(\nu )
	\,.
}
One can confirm that the long wavelength asymptotes of the renormalized bias functions reduce to the bias defined above, namely
$c_n^{\rm L}(|{\bf k}_i|\to 0)\to b_n^{\rm L}$.
In this paper, we employ the Sheth-Tormen fitting formula~\citep{Sheth:1999su} for the universal mass function, which is given by
\al{
	f_{\rm MF}(\nu )=A\sqrt{\frac{2q}{\pi}}\left( 1+(q\nu^2)^{-p}\right)\nu e^{-q\nu^2/2}
	\,,\label{eq:ST mass function}
}
with $A=[1+\Gamma(1/2-p)/\sqrt{\pi}2^p]^{-1}$\,, $p=0.3$\,, and $q=0.707$.
When the power spectrum and bispectrum of galaxy clustering are considered, the detailed amplitude and shape of 
scale-dependent biases depend on the first three functions of the renormalized bias functions, $c_1^{\rm L}({\bf k})$, $c_2^{\rm L}({\bf k}_1,{\bf k}_2)$, and $c_3^{\rm L}({\bf k}_1,{\bf k}_2,{\bf k}_3)$. 
In particular, the coefficients of the bias parameters $A_1$, $A_2$, and $A_3$ are explicitly written in terms of the parameters
of the Sheth-Tormen mass functions as
\al{
	&&A_1(M)=q\nu^2+\frac{2p}{1+(q\nu^2)^p}
	\,,\label{eq:A1}\\
	&&A_2(M)=q\nu^2\delta_{\rm c}b_1^{\rm L}(M)
	            +\frac{2p(q\nu^2+2p+1)}{1+(q\nu^2)^p}
	\,,\\
	&&A_3(M)=q\nu^2\delta_{\rm c}^2b_2^{\rm L}(M)
	            +\frac{2p[2p(2q\nu^2+2p+3)+q^2\nu^4+q\nu^2+2]}{1+(q\nu^2)^p}
	 \,.\label{eq:A3}
}
We also write down the Lagrangian bias functions as 
\al{
	&&b_1^{\rm L}(M)=\frac{1}{\delta_{\rm c}}
	    \left(
	        q\nu^2-1+\frac{2p}{1+(q\nu^2)^p}
	    \right)
	\,,\\
	&&b_2^{\rm L}(M)=\frac{1}{\delta_{\rm c}^2}
	    \left(
	        q^2\nu^4-3q\nu^2+\frac{2p(2q\nu^2+2p-1)}{1+(q\nu^2)^p}
        \right)
	\,,\\
	&&b_3^{\rm L}(M)=\frac{1}{\delta_{\rm c}^3}
	    \left(
            q^3\nu^6-6q^2\nu^4+3q\nu^2
            +\frac{2p[2p(3q\nu^2+2p)+3q^2\nu^4-6q\nu^2-1]}{1+(q\nu^2)^p}
	    \right) \,.
}

\end{document}